\documentclass[11pt]{article}
\usepackage[a4paper, margin=1in]{geometry}

\usepackage[usenames]{color}
\usepackage[figuresright]{rotating}
\usepackage{boxedminipage}
\usepackage[colorlinks,citecolor=blue,urlcolor=blue]{hyperref}
\usepackage{epstopdf}
\usepackage{natbib}
\usepackage{multirow}
\usepackage{enumerate}
\usepackage{enumitem}
\usepackage{verbatim}
\usepackage{bbm}
\usepackage[dvipsnames]{xcolor}
\usepackage{bm}
\usepackage{mathrsfs,color,dsfont}
\usepackage{algorithm}
\usepackage{algorithmic}
\usepackage{extarrows}
\usepackage{subcaption}
\usepackage{booktabs}
\usepackage{array} 
\usepackage{makecell}

\usepackage{tikz}
\usetikzlibrary{arrows}

\usepackage{amsthm,amsmath,amssymb,amsfonts, amsbsy, mathtools, epsfig}

\newcommand{\blind}{1}

\renewcommand{\baselinestretch}{1.31}
\newcommand{\linsp}{\renewcommand{\baselinestretch}{1.31}}
\newcommand{\linsps}{\renewcommand{\baselinestretch}{1.3}}

\newcommand{\vect}[1]{\boldsymbol{#1}}				

\newcommand{\tp}{{\top}}						
\newcommand{\diag}{{\rm diag}}					
\newcommand{\med}[1]{\mathop{\rm{med}}_{#1}}	
\newcommand{\diff}{\mathrm{d}}					
\newcommand{\argmin}[1]{\mathop{\rm{argmin}}_{#1}}	
\newcommand{\argmax}[1]{\mathop{\rm{argmax}}_{#1}}	
											
\newcommand*\widebar[1]{\,						
	\hbox{%
   		\kern0.01em%
		\vbox{%
			\hrule height 0.5pt		
			\kern0.33ex			
			\hbox{%
				\kern-0.1em		
				\ensuremath{#1}%
				\kern-0.1em		
			}%
		}%
	}%
\,}%
\let\hat\widehat
\let\tilde\widetilde
\let\bar\widebar

\newcommand{\mcB}{{\mathcal B}}					

\newcommand{\mcF}{{\mathcal F}}

\newcommand{\mcH}{{\mathcal H}}					

\newcommand{\mcL}{{\mathcal L}}					

\newcommand{\mcN}{{\mathcal N}}					

\newcommand{\mcQ}{{\mathcal Q}}

\newcommand{\mcS}{{\mathcal S}}

\newcommand{\mcX}{{\mathcal X}}

\newcommand{\mbB}{{\mathbb B}}					

\newcommand{\mbE}{{\mathbb E}}					

\newcommand{\mbI}{{\mathbb I}}					

\newcommand{\mbP}{{\mathbb P}}					
\newcommand{\mbR}{{\mathbb R}}					
\newcommand{\mbS}{{\mathbb S}}					

\newcommand{\mbX}{{\mathbb X}}

\DeclareMathAlphabet\EuScriptBF{U}{eus}{b}{n}

\newcommand{\kl}{\mathrm{KL}}
\newcommand{\elbo}{\mathrm{ELBO}}
\newcommand{\mvb}{\mathrm{M}^3\mathrm{VB}}

\newcommand{\quan}{\mathrm{Quan}}	

\definecolor{DSgray}{cmyk}{0,1,0,0}

\definecolor{scolor}{cmyk}{0.5,2,0,0}

\newtheorem{theorem}{Theorem}
\newtheorem{corollary}[theorem]{Corollary}
\newtheorem{proposition}[theorem]{Proposition}
\newtheorem{remark}{Remark}
\newtheorem{example}{Example}
\newtheorem{condition}{Condition}

\begin{document}


\linsps

\if1\blind
{
	\title{\bf Robust Variational Bayes by Min-Max Median Aggregation}
		 \author{Jiawei Yan\thanks{School of Mathematical Sciences, Shanghai Jiao Tong University},\quad Ju Liu\thanks{School of Mathematics, Shanghai University of Finance and Economics},\quad Weidong Liu\thanks{School of Mathematical Sciences and MoE Key Lab of Artificial Intelligence, Shanghai Jiao Tong University},\quad and Jiyuan Tu\thanks{School of Statistics and Data Science, Shanghai University of Finance and Economics }
		 \date{}
}		\maketitle
} \fi

\if0\blind
{
	\bigskip
	\bigskip
	\bigskip
	\begin{center}
		{\LARGE\bf Generalized Rank Regression}
	\end{center}
	\medskip
} \fi


\bigskip
\begin{abstract}
We propose a robust and scalable variational Bayes (VB) framework designed to effectively handle contamination and outliers in dataset. Our approach partitions the data into $m$ disjoint subsets and formulates a joint optimization problem based on robust aggregation principles. A key insight is that the full posterior distribution is equivalent to the minimizer of the mean Kullback-Leibler (KL) divergence from the $m$-powered local posterior distributions. To enhance robustness, we replace the mean KL divergence with a min-max median formulation. The min-max formulation not only ensures consistency between the KL minimizer and the Evidence Lower Bound (ELBO) maximizer but also facilitates the establishment of improved statistical rates for the mean of variational posterior. We observe a notable discrepancy in the $m$-powered marginal log likelihood function contingent on the presence of local latent variables. To address this, we treat these two scenarios separately to guarantee the consistency of the aggregated variational posterior. Specifically, when local latent variables are present, we introduce an aggregate-and-rescale strategy. Theoretically, we provide a non-asymptotic analysis of our proposed posterior, incorporating a refined analysis of Bernstein-von Mises (BvM) theorem to accommodate a diverging number of subsets $m$. Our findings indicate that the two-stage approach yields a smaller approximation error compared to directly aggregating the $m$-powered local posteriors. Furthermore, we establish a nearly optimal statistical rate for the mean of the proposed posterior, advancing existing theories related to min-max median estimators. The efficacy of our method is demonstrated through extensive simulation studies. 
\end{abstract}


\linsp


\section{Introduction}

Bayesian inference is a fundamental approach in statistics and machine learning, offering a rigorous framework for probabilistic modeling. Typically, this involves specifying a probability model that captures the relationship between latent variables and observed data, incorporating prior information about the latent variables and a likelihood function. The goal is to infer the posterior distribution of the latent variables given the observations. To exactly approximate the posterior distribution, Markov chain Monte Carlo (MCMC) method \citep{hastings.1970bmk, Gelfand_smith.1990jasa} are commonly employed. These methods construct a Markov chain whose stationary distribution is the target posterior, allowing for sampling-based approximation. However, MCMC methods often suffer from computational intractability and autocorrelation among generated samples, especially in high-dimensional settings. As an alternative, variational Bayes (VB) methods \citep{jordan.1999ml, bishop.2006} have emerged as a popular approach.

VB methods approximate the posterior distribution by a simpler, tractable distribution, transforming the inference problem into an optimization task. This approach offers improved computational efficiency and scalability, making it suitable for large-scale applications. VB methods have been successfully applied to a variety of probabilistic models. Notable examples include the Gaussian mixture model \citep{robert_etal.1998tpami}, latent Dirichlet allocation \citep{blei.2003jmlr}, and the stochastic block model \citep{hofman_wiggins.2008prl}. For comprehensive overviews of VB techniques and their applications, readers are referred to \citep{Blei_etal.2017jasa, zhang_etal.2019tpami}.

Under certain technical conditions, such as sufficient samples size and data independence, variational posterior distributions have been shown to exhibit desirable theoretical properties \citep{wang_blei.2019jasa}. The theoretical foundations of Bayesian learning trace back to the seminal work of \cite{doob1949application}, with comprehensive treatments provided by \cite{LeCam.2000, vanDerVaart.2000}. Central to this framework is the Bernstein-von Mises (BvM) theorem, which asserts that, under certain regularity conditions, the posterior distribution converges to a normal distribution as the sample size approaches infinity. Extension of the BvM theorem have been explored in various contexts such as semi-parametric learning \citep{bickel_kleijn.2012aos}, posterior with misspecification \citep{kleijn_vaart.2012ejs}, general Bayes \citep{bissiri_etal.2016jrssb, miller.2021jmlr, knoblauch_etal.2022jmlr}, and non-asymptotic analysis accommodating growing parameter dimensions \citep{spokoiny2014arXiv}. In the realm of variational Bayes, \citet{wang_blei.2019jasa} initiated the study of BvM-type results for VB, and was further extended to misspecified setting \citep{wang_blei.2019nips}, variational posterior \citep{zhang_gao.2020aos}, and their applications in model selection \citep{zhang_yang.2024jrssb} and multiplier bootstrap techniques \citep{han_yang.2019arXiv}. 

Outliers and data corruption are ubiquitous in modern data analysis. Traditional VB methods can be severely compromised by even a single outlier, highlighting the need for robust VB approaches. Despite its importance, robust Bayesian learning remains a relatively underexplored area. For full Bayesian posteriors, \cite{minsker_etal.2014, minsker_etal.2017jmlr} proposed aggregating local posterior distributions using the geometric median, enhancing robustness against outliers. This concept was extended to the variational Bayes framework by \cite{padilla_etal.2025arXiv}, though without incorporating local latent variables. These approaches typically depend on defining a metric over the space of probability measure, which is often computationally intensive in practice. Moreover, their algorithms require model-specific designs, limiting their general applicability. In a different vein, \cite{minsker_yao.2025ml} introduced a robust Bayesian learning method based on the median-of-means principle, applying this framework to log-likelihood functions. While this approach enhances robustness, the resulting posterior distribution is challenging to sample from and does not readily integrate into the variational Bayes framework.

In this paper, we address the problem of robust variational Bayes from a novel perspective. Following a strategy similar to that of \cite{padilla_etal.2025arXiv}, we partition the dataset into $m$ disjoint subsets and aim to aggregate them in a robust manner. Our key observation is that the full posterior distribution can be characterized as the minimizer of the mean KL divergence from the $m$-powered local posterior distributions (see Proposition \ref{prop:mean_KL} for further details). Building on the insights of \cite{ lecue_lerasle.2020aos}, we propose a min-max median aggregation framework over the individual posterior distributions, called the $\mvb$ method. In contrast to previous approaches such as \cite{minsker_etal.2014, minsker_etal.2017jmlr, padilla_etal.2025arXiv}, which require explicit $m$-powered local (variational) posterior distributions, our method formulates the aggregation as a joint optimization problem that only involves the likelihood functions, without dependence on the normalizing constants. The resulting optimization problem can be efficiently solved using a variant of the coordinate ascent algorithm \citep{bishop.2006, wright.2015mp}, leading to substantial improvements in both computational efficiency and practical applicability compared to existing methods.

In several important applications, such as Bayesian mixture model and stochastic block model, local latent variables appear explicitly in the likelihood functions. Intriguingly, we observe that when such local latent variables are present, the marginal form of the $m$-powered likelihood function is neither identical nor proportional to the $m$-th power of the marginal likelihood function. This subtle discrepancy can lead to inconsistency in the posterior mean. To address this issue, we propose treating models with and without local latent variables separately. Specifically, in the presence of local latent variables, we avoid taking the $m$-th power of the local posterior distributions and instead apply the min-max median aggregation directly to the unpowered posteriors. To preserve statistical efficiency under this revised scheme, we introduce a rescaling factor of $1/m$. Theoretically, we establish that the proposed aggregate-and-rescale strategy yields a smaller approximation error compared to the direct aggregation of $m$-powered local posteriors.

From a theoretical perspective, existing VB analysis typically rely on Markov's inequality to control tail probabilities. While this approach yields coarse bounds, it is inadequate for uniformly controlling the asymptotic error across all local distributions, particularly when the number of subsets $m$ increases with the sample size. To accommodate the regime of diverging $m$, we reestablish the VB theoretical framework using sharper concentration inequalities. Our analysis circumvents the commonly used, yet indirect testability condition \citep{kleijn_vaart.2012ejs, wang_blei.2019jasa, zhang_yang.2024jrssb}, and is sufficiently general to encompass the generalized Bayesian framework \citep{bissiri_etal.2016jrssb, miller.2021jmlr, knoblauch_etal.2022jmlr} as well as models with potential misspecification \citep{kleijn_vaart.2012ejs, wang_blei.2019nips, zhang_yang.2024jrssb}. Under this refined theoretical foundation, we establish a Bernstein-von Mises (BvM) theorem for our proposed $\mvb$ method, treating models with and without local latent variables separately. Furthermore, we theoretically demonstrate that our method is robust to $\alpha_n$-level contamination, for any $\alpha_n\in[0,1/2-\delta)$ with $\delta\in[0,1/2)$.

We have analyzed the statistical convergence rate of the mean of the $\mvb$ distribution. Our findings indicate that this rate is asymptotically equivalent to that of the min-max median of the log likelihood function, as established by \cite{lecue_lerasle.2020aos}. Specifically, we demonstrate that the mean achieves a convergence rate of $O_{\mbP}(\alpha_n/\sqrt{n}+1/\sqrt{mn}+1/n^{3/4})$, which is nearly optimal when $m=O(\sqrt{n})$ or $n=O(\alpha_n^{-4})$. This rate improves upon the $O_{\mbP}(\sqrt{\alpha_n/n})$ rate reported in \cite{lecue_lerasle.2020aos}, which deserves independent interest. The key insight underpinning this improvement is that the min-max formulation effectively cancels out the quadratic term and symmetrizes the distribution of each term in the median. This symmetrization allows us to exploit the coincidence of the median and mean in symmetric distributions, as discussed in \cite{minsker.2019ejs}, thereby attaining a superior convergence rate.

In summary, our contribution is three fold: Firstly, we propose a min-max median aggregation approach for variational Bayes, which is much computationally cheaper than existing robust Bayesian methods; Secondly, we have discover an interesting discrepancy for variational Bayesian with or without the presence of local latent variables, and therefore propose to address them separately; Thirdly, we have established non-asymptotic analysis for our proposed approach, our theory is general enough to encompass more settings, and is sharp enough to obtain precise statistical rate.


\subsection{Paper Organization and Notations}
The remainder of this paper is organized as follows. In Section \ref{sec:m3vb_org}, we introduce the min-max median variational Bayes ($\mvb$) method in the absence of local latent variables. Section \ref{sec:theory_M3VB} presents the theoretical foundations of our approach. In Section \ref{sec:med_elbo}, we extend the $\mvb$ framework to accommodate models with local latent variables. Section \ref{sec:alg} details the algorithmic implementation of our method and provides illustrative examples. In Section \ref{sec:sim}, we conduct numerical experiments to demonstrate the effectiveness of the proposed approach. Concluding remarks are offered in Section \ref{sec:conclude}. All proofs are deferred to the Appendix.

For a vector $\vect{v}=(v_1,...,v_p)^{\tp}$, we denote the $\ell_q$ norm by $|\vect{v}|_q=(\sum_{l=1}^p|v_l|^q)^{1/q}$ and the $\ell_{\infty}$ norm by $|\vect{v}|_{\infty}=\max_{1\leq l\leq p}|v_l|$. We use an uppercase letter $P$ to denote a probability distribution, and the corresponding lowercase letter $p$ to denote its density function. The mean and covariance of the distribution $P$ are denoted by $\vect{\mu}_P$ and $\vect{\Sigma}_P$, respectively. For a symmetric matrix $\vect{A}$, we denote its largest and smallest eigenvalues by $\Lambda_{\max}(\vect{A})$ and $\Lambda_{\min}(\vect{A})$, respectively. The identity matrix of dimension $p$ is denoted by $\mbI_p$. For simplicity, we use $\mbS_r^{p-1}(\vect{v})$ and $\mbB_r^p(\vect{v})$ to represent the sphere and the closed ball of radius $r$ centered at $\vect{v}\in\mbR^p$, respectively. Given a sequence of real numbers $\{x_i\}_{i=1}^n$, we denote its median by $\med{1\leq i\leq n}\{x_i\}$ and its $\tau$-th quantile by $\quan_{\tau}(x_i|1\leq j\leq n)$, where $\tau\in(0,1)$ is the quantile level. We use $\otimes$ to denote the product of probability measures. The parameter space, covariate space, and local latent variable space are denoted by $\Theta$, $\mcX$ and $\mcS$, respectively. Unless otherwise specified, generic constant are assumed to be independent of $m,n$ and $p$.


\section{Min-Max Median Variational Bayes}	\label{sec:m3vb_org}

In this section, we introduce the core concept of our proposed min-max median variational Bayes ($\mvb$) method. To provide a solid foundation, we begin by reviewing essential preliminaries of variational Bayes.


\subsection{Preliminaries of Variational Bayes}	\label{sec:pre_VB}

Let $\mbX = \{\vect{X}_1,...,\vect{X}_N\}$ denotes the set of $N$  i.i.d. random variables drawn from an unknown data-generating distribution $P_0$. We consider a parametric family of distributions $\{P_{\vect{\theta}}|\vect{\theta}\in\Theta\}$, which may not necessarily include the true distribution $P_0$. Within the Bayesian learning framework, our objective is to approximate the posterior distribution $P(\vect{\theta}|\mbX)$ of the parameter $\vect{\theta}\in\Theta$, given the dataset $\mbX$ and a prior distribution $\pi(\vect{\theta})$. The posterior density is given by
\begin{equation}	\label{eq:post_def}
	p(\vect{\theta}|\mbX) = \frac{\pi(\vect{\theta})p(\mbX|\vect{\theta})}{p(\mbX)} =\frac{\pi(\vect{\theta})\prod_{i=1}^Np(\vect{X}_i|\vect{\theta})}{p(\mbX)},\quad\text{where }p(\mbX) = \int_{\Theta}p(\mbX|\vect{\theta})\pi(\vect{\theta})\diff\vect{\theta}.
\end{equation}
Since the posterior distribution in \eqref{eq:post_def} is generally computationally intensive to sample from \citep{hastings.1970bmk, Gelfand_smith.1990jasa}, variational Bayes (VB) has been proposed as an efficient alternative for approximating the posterior using a more tractable distribution from a predefined distribution family $\mcQ$. Specifically, VB seeks a distribution $\hat{Q}_{\vect{\theta}}$ that minimizes the KL divergence
\begin{equation}	\label{eq:KL_min}
	\hat{Q}_{\vect{\theta}} = \argmin{Q_{\vect{\theta}}\in\mcQ}\kl\big(Q_{\vect{\theta}}\big\|P(\vect{\theta}|\mbX)\big).
\end{equation}
However, directly minimizing the KL divergence is challenging due to the intractability of the true posterior $P(\vect{\theta}|\mbX)$. To circumvent this, we observe that:
\begin{align*}
	\kl\big(Q_{\vect{\theta}}\big\|P(\vect{\theta}|\mbX)\big) =& \int_{\Theta} q_{\vect{\theta}}(\vect{\theta})\log\Big(\frac{q_{\vect{\theta}}(\vect{\theta})}{p(\vect{\theta}|\mbX)}\Big)\diff\vect{\theta}\\
	=& \log p(\mbX) - \int_{\Theta} q_{\vect{\theta}}(\vect{\theta})\log\Big(\frac{p(\vect{\theta},\mbX)}{q_{\vect{\theta}}(\vect{\theta})}\Big)\diff\vect{\theta},
\end{align*}
where the first term $\log p(\mbX)$ is independent of $q_{\vect{\theta}}$ and can thus be omitted during optimization. Consequently, the minimization problem in \eqref{eq:KL_min} is equivalent to maximizing the evidence lower bound ($\elbo$):
\begin{equation}	\label{eq:elbo_max}
	\hat{Q}_{\vect{\theta}} = \argmax{Q_{\vect{\theta}}\in\mcQ}\int_{\Theta}q_{\vect{\theta}}(\vect{\theta})\log\Big(\frac{p(\vect{\theta},\mbX)}{q_{\vect{\theta}}(\vect{\theta})}\Big)\diff\vect{\theta} =:\argmax{Q_{\vect{\theta}}\in\mcQ}\elbo(Q_{\vect{\theta}}\| P(\vect{\theta},\mbX)).
\end{equation}
The  $\elbo$ serves as a lower bound on the marginal log-likelihood $\log p(\mbX)$, owing to the non-negativity of the KL divergence.

In practical applications, a commonly adopted choice for the variational family $\mcQ$ is the mean-field family, which comprises all factorized densities of the form $q(\vect{\theta})=\prod_{l=1}^pq_l(\theta_l)$ for $\vect{\theta}\in\Theta$. Alternative structures, such as the block mean-field family \citep{carbonetto_stephens.2012ba}, have also been proposed to capture dependencies among subsets of parameters.

In certain models, such as the Gaussian mixture model, the introduction of local latent variable significantly facilitates the evaluation of the likelihood function. In our methodology, posterior approximations with and without local latent variables necessitate distince treatments. Consequently, we defer the detailed formulation of variational Bayes method incorporating local latent variables to Section \ref{sec:med_elbo}.


\subsection{Min-Max Median Aggregation of Distributions}	\label{sec:mean_dist}

Suppose we evenly divide the dataset $\mbX=\{\vect{X}_1,...,\vect{X}_N\}$ into $m$ subsets $\mbX_j=\{\vect{X}_i|i\in\mcH_j\}$, where $\mcH_j$, for $1\leq j\leq m$, denotes the index set corresponding to the $j$-th subset. We assume that an $\alpha_n$ fraction of these subsets may be arbitrarily corrupted, and denote the set of corrupted indices by $\mcB\subseteq\{1,...,m\}$, with $|\mcB|=\lfloor \alpha_nm\rfloor$. Our goal is to construct local posterior distributions $P_j$ based on each sub-dataset $\mbX_j$ and robustly aggregate them to approximate the full posterior distribution $P(\vect{\theta}|\mbX)$ as defined in \eqref{eq:post_def}. A naive approach is to compute the local posterior
\begin{equation*}
	p(\vect{\theta}|\mbX_j) = \frac{\pi(\vect{\theta})\prod_{i\in\mcH_j}p(\vect{X}_i|\vect{\theta})}{\int_{\Theta}\pi(\vect{\theta})\prod_{i\in\mcH_j}p(\vect{X}_i|\vect{\theta})\diff\vect{\theta}},
\end{equation*}
and then aggregate them using the geometric median under a suitable metric on the space of probability measures \citep{minsker_etal.2014, minsker_etal.2017jmlr, padilla_etal.2025arXiv}. However, as noted in prior work \citep{minsker_etal.2014, minsker_etal.2017jmlr, li_etal.2017bmk, srivastava_etal.2018jmlr}, this approach suffers from a scale mismatch in covariance. Let $\vect{H}$ denote the Fisher information matrix of the model. By the Bernstein-von Mises theorem, the local posterior has asymptotic covariance $\frac{m}{n}\vect{H}$, whereas the full posterior has asymptotic covariance $\frac{1}{n}\vect{H}$, which is much smaller. To resolve this discrepancy, the aforementioned studies proposed aggregating the $m$-powered local posterior distribution $\tilde{P^m}(\vect{\theta}|\mbX_j)$, defined via the density 
	\begin{equation}	\label{eq:mpower_post}
		\tilde{p^m}(\vect{\theta}|\mbX_j) = \frac{\pi(\vect{\theta})\prod_{i\in\mcH_j}p^m(\vect{X}_i|\vect{\theta})}{\int_{\Theta} \pi(\vect{\theta})\prod_{i\in\mcH_j}p^m(\vect{X}_i|\vect{\theta})\diff\vect{\theta}}\propto \pi(\vect{\theta})\prod_{i\in\mcH_j}p^m(\vect{X}_i|\vect{\theta}).
	\end{equation}
Indeed, we have the following result.

\begin{proposition}	\label{prop:mean_KL}
	Let $\tilde{P^m}(\vect{\theta}|\mbX_j)$ be defined as in equation \eqref{eq:mpower_post}. Then, the full posterior distribution $P(\vect{\theta}|\mbX)$ satisfies
	\begin{equation}	\label{eq:KL_mean}
	\begin{aligned}
		P(\vect{\theta}|\mbX) =&\argmin{Q_{\vect{\theta}}}\Big\{\frac{1}{m}\sum_{j=1}^{m}\kl \big(Q_{\vect{\theta}}\|\tilde{P^m}(\vect{\theta}|\mbX_j)\big)\Big\}\\
		=&\argmax{Q_{\vect{\theta}}}\Big\{\frac{1}{m}\sum_{j=1}^{m}\elbo\big(Q_{\vect{\theta}}\|\tilde{P^m}(\vect{\theta},\mbX_j)\big)\Big\},
	\end{aligned}
	\end{equation}
	where $\tilde{P^m}(\vect{\theta},\mbX_j)$ has the density function
	\begin{align*}
		\tilde{p^m}(\vect{\theta},\mbX_j) = \frac{\pi(\vect{\theta})p^m(\mbX_j|\vect{\theta})}{\int_{\Theta} \pi(\vect{\theta})p^m(\mbX_j|\vect{\theta})\diff\vect{\theta}\diff\mbX_j}.
	\end{align*}
\end{proposition}

The above proposition offers another compelling justification for adopting the $m$-powered formulation of local posterior distributions. Specifically, it shows that the full posterior $P(\vect{\theta}|\mbX)$ coincides with the minimizer of the average KL divergence from the $m$-powered local posteriors. However, this naive aggregation scheme is inherently non-robust. In particular, even a single outlying subset can cause the aggregated distribution to concentrate arbitrarily far from the true parameter value. Motivated by Proposition \ref{prop:mean_KL} and recent developments in robust estimation based on medians \citep{minsker.2019ejs, lecue_lerasle.2019spa, lecue_etal.2020ml}, a natural remedy is to replace the mean KL divergence with the median. This leads to the following robust aggregation formulation:
\begin{equation}	\label{eq:naive_KLVB_median}
	\hat{Q}_{\vect{\theta}} =\argmin{Q_{\vect{\theta}}\in\mcQ}\med{1\leq j\leq m}\Big\{\kl \big(Q_{\vect{\theta}}(\vect{\theta})\|\tilde{P^m}(\vect{\theta}|\mbX_j)\big)\Big\}.
\end{equation}
Here the variational distribution $Q_{\vect{\theta}}$ is restricted to a variational family $\mcQ$, leading naturally to a robust variational Bayes (VB) method. While this approach is designed to yield a robust aggregation, it suffers from a key drawback arising from the nonlinearity of the median operator. In particular, observe that
\begin{align*}
	&\argmin{Q_{\vect{\theta}}\in\mcQ}\med{1\leq j\leq m}\Big\{\kl \big(Q_{\vect{\theta}}(\vect{\theta})\|\tilde{P^m}(\vect{\theta}|\mbX_j)\big)\Big\}\\
	=&\argmin{Q_{\vect{\theta}}\in\mcQ}\med{1\leq j\leq m}\Big\{\tilde{P^m}(\mbX_j) - \elbo \big(Q_{\vect{\theta}}(\vect{\theta})\|\tilde{P^m}(\vect{\theta},\mbX_j)\big)\Big\}\\
	\neq&\argmax{Q_{\vect{\theta}}\in\mcQ}\med{1\leq j\leq m}\Big\{ \elbo \big(Q_{\vect{\theta}}(\vect{\theta})\|\tilde{P^m}(\vect{\theta},\mbX_j)\big)\Big\},
\end{align*}
where the last inequality is due to $\tilde{P^m}(\mbX_j)$ differs across $j$. This discrepancy implies that the minimizer of the median KL divergence does not, in general, coincide with the maximizer of the median ELBO. Such a mismatch is undesirable and introduces complications in theoretical analysis. To address this issue, we draw inspiration from the recent min-max median frameworks developed in \cite{lecue_lerasle.2019spa, lecue_lerasle.2020aos} for empirical risk minimization, and propose the following min-max formulation:

\begin{align*}
	\hat{F}(\vect{\theta}) =&\argmin{F\in\mcQ}\max_{G\in\mcQ}\med{1\leq j\leq m}\Big\{\kl \big(F(\vect{\theta})\|\tilde{P^m}(\vect{\theta}|\mbX_j)\big) - \kl \big(G(\vect{\theta})\|\tilde{P^m}(\vect{\theta}|\mbX_j)\big)\Big\}\\
	=&\argmin{F\in\mcQ}\max_{G\in\mcQ}\med{1\leq j\leq m}\Big\{\elbo \big(G(\vect{\theta})\|\tilde{P^m}(\vect{\theta},\mbX_j)\big)-\elbo \big(F(\vect{\theta})\|\tilde{P^m}(\vect{\theta},\mbX_j)\big) \Big\}.\stepcounter{equation}\tag{\theequation}\label{eq:minmax_median}
\end{align*}
The derivation above reveals a key benefit of the min-max formulation: the marginal likelihood terms $\tilde{P^m}(\mbX_j)$ cancel out, ensuring that the KL minimizer aligns with the ELBO maximizer. This alignment simplifies analysis and yields a more coherent objective. A further, more subtle advantage of this formulation will be discussed in Section \ref{sec:mu_rate}.  Hereafter, we refer to the proposed method in \eqref{eq:minmax_median} as the min-max median variational Bayes ($\mvb$) method.

\begin{remark}
	Most existing approaches, such as those in \cite{srivastava_etal.2015aistats, li_etal.2017bmk, srivastava_etal.2018jmlr, minsker_etal.2017jmlr, padilla_etal.2025arXiv}, first construct (variational) posterior distribution using local data subsets, and then aggregate these distributions based on certain metrics defined over the space of probability measures, such as the Wasserstein distance or norms induced by reproducing kernel Hilbert space (RKHS). However, generating these local distributions can be computationally expensive, and the subsequent aggregation process-especially when constrained to a specific family of distributions-is often intractable and instance-specific. Furthermore, because the KL divergence does not constitute a metric on the space of probability distributions, our method fundamentally differs from these distance-based approaches, thereby necessitating a distinct theoretical framework. In contrast, our method in equation \eqref{eq:minmax_median} directly formulates a joint optimization problem that avoids the explicit computation of local posterior distributions. As we elaborate in Section \ref{sec:alg}, our approach can be efficiently implemented via a straightforward modification of classical algorithms, such as coordinate ascent \citep{bishop.2006, wright.2015mp}.
\end{remark}

\begin{remark}
	A closely related approach is proposed in \cite{minsker_yao.2025ml}, where the authors consider computing the local log likelihood difference
	\begin{equation}	\label{eq:minsker_yao}
		L(\vect{\theta},\mbX_j) = \sum_{i\in\mcH_j}\big\{\log(p(\vect{\theta}|\vect{X}_i)) - \log(p(\vect{\theta}'|\vect{X}_i))\big\}
	\end{equation}
	based on the local dataset indexed by $\mcH_j$, where $\vect{\theta}'$ is a fixed parameter. Then they aggregate the local quantities into a global statistic $\hat{L}(\vect{\theta})$ using a Huber-type robust aggregator, and subsequently sample from the distribution $\exp(\hat{L}(\vect{\theta}))$ via Monte-Carlo methods. While both their approach and ours share a loss-difference formulation, there are several fundamental differences. First, because $\vect{\theta}'$ is fixed in \eqref{eq:minsker_yao}, the aggregated loss $\hat{L}(\vect{\theta})$ depends on a single parameter and does not take a min-max form, in contrast to our formulation. Second, their use of a Huber-type aggregator resembles a truncated mean, which behaves similarly to the arithmetic average of local log likelihood functions. This yields different theoretical properties compared to our median-based aggregation. Most importantly, their method aggregates local log likelihood functions, whereas our approach focuses on robust aggregation of KL divergences. Furthermore, their procedure is computationally intensive due to the need for sampling and is not readily extendable to variational Bayes framework.
\end{remark}


\section{Theory of Min-Max Median Variational Bayes}	\label{sec:theory_M3VB}

In this section, we present the theoretical underpinnings of the proposed $\mvb$ estimator. Prior to detailing the main results, we introduce several technical conditions essential for our analysis. Throughout this paper, we define the log-likelihood function as $\ell(\vect{\theta},\vect{X}) = \log p(\vect{X}|\vect{\theta})$, identify the true parameter as $\vect{\theta}^* =\argmax{\vect{\theta}}\mbE[\ell(\vect{\theta},\vect{X})]$, and denote the Fisher information matrix as $\vect{H} = -\mbE[\nabla^2\ell(\vect{\theta}^*,\vect{X})]$. For analytical tractability, we constrain the variational family $\mcQ$ to consist of factorized Gaussian distributions, a subclass within the mean-field variational family. This assumption facilitates our theoretical development by obviating the need to establish the sub-Gaussianity of the estimator $\hat{F}_{\vect{\theta}}$, a property whose verification typically depends on the convexity of the objective function \citep{han_yang.2019arXiv, zhang_yang.2024jrssb}, which may not hold in our framework. We denote $\mcQ^p$ as the family of factorized Gaussian distributions with parameter dimension $p$.

\begin{condition}	\label{cond:thick_prior}
	The prior distribution $\pi(\vect{\theta})$ satisfies $\pi(\vect{\theta}^*)>0$, and there exist constants $\eta_1,M_1>0$ such that
	\begin{equation*}
		|\log \pi(\vect{\theta}) - \log \pi(\vect{\theta}^*)|\leq M_1(1+|\vect{\theta} - \vect{\theta}^*|_2^{\eta_1}),\quad \theta\in\Theta.
	\end{equation*}
\end{condition}

\begin{condition}	\label{cond:convex}
	The parameter space $\Theta$ is uniformly bounded. Moreover, there exists a constant $c_l>0$ such that
	\begin{align*}
		\mbE[\ell(\vect{\theta}^*,\vect{X})]-\mbE[\ell(\vect{\theta},\vect{X})]\geq  c_l|\vect{\theta} - \vect{\theta}^*|_2^2.
	\end{align*}
	Additionally, there exists a constant $\rho\in(0,1]$ such that
	\begin{equation*}
		\rho^{-1}\geq \Lambda_{\max}(\vect{H})\geq \Lambda_{\min}(\vect{H})\geq \rho.
	\end{equation*}
\end{condition}

\begin{condition}	\label{cond:subgauss_Grad}
	There exist constants $\eta_2,M_2>0$ such that
	\begin{align*}
		\sup_{\vect{v}\in\mbS^{p-1}}\sup_{\vect{\theta}\in\Theta}\mbE\Big[\exp\Big\{\eta_2\big|\vect{v}^{\tp}\nabla\ell(\vect{\theta},\vect{X})\big|\Big\}\Big]\leq M_2.
	\end{align*}
\end{condition}

\begin{condition}	\label{cond:subgauss_Hess}
	There exist constants $\eta_3,M_3>0$ such that
	\begin{align*}
		\sup_{\vect{v}\in\mbS^{p-1}}\sup_{\vect{\theta}\in\Theta}\mbE\Big[\exp\Big\{\eta_3\vect{v}^{\tp}\nabla^2\ell(\vect{\theta},\vect{X})\vect{v}\Big\}\Big]\leq M_3.
	\end{align*}
\end{condition}

\begin{condition}	\label{cond:L_smooth}
	The log-likelihood function $\ell(\vect{\theta},\vect{X})$ is three times differentiable. Specifically, there exists a constant $L_1>0$ such that
	\begin{align*}
		\Big|\frac{\partial^3}{\partial\theta_i\partial\theta_j\partial\theta_k}\ell(\vect{\theta},\vect{X})\Big|\leq L_1,\quad 1\leq i,j,k\leq p.
	\end{align*}
\end{condition}

Condition \ref{cond:thick_prior} imposes a mild regularity requirement on the prior distribution $\pi(\vect{\theta})$, ensuring sufficient control over its tail behavior. Condition \ref{cond:convex} guarantees a global quadratic growth of the log-likelihood function, which is essential for controlling the tail behavior of the posterior distribution. This assumption is also adopted in \cite{spokoiny2014arXiv}, albeit in a relaxed form where the constant $c_l$ is replaced by a positive, decreasing function of the distance $|\vect{\theta} - \vect{\theta}^*|_2$. While this relaxation broadens the class of admissible models, it complicates the theoretical analysis. Our framework can be extended to accommodate such a relaxation; however, we omit the details for brevity. Condition \ref{cond:convex} also assumes that the parameter space $\Theta$ is uniformly bounded, an assumption that is implicitly made in prior works such as \cite{ghosal_etal.2000aos, han_yang.2019arXiv, zhang_yang.2024jrssb} (see Section 5 of \cite{ghosal_etal.2000aos} for more details). Conditions \ref{cond:subgauss_Grad} and \ref{cond:subgauss_Hess} assume that the gradient and Hessian of the log-likelihood function exhibit sub-exponential tail, which aligns with assumptions commonly made in the literature, including \cite{han_yang.2019arXiv, zhang_yang.2024jrssb}. Condition \ref{cond:L_smooth} requires the third-order derivatives of the log-likelihood function to be uniformly bounded. This assumption is slightly stronger than those in \cite{han_yang.2019arXiv, zhang_yang.2024jrssb}, and is adopted here to simplify the theoretical analysis. Nevertheless, our analysis could be extended to the weaker conditions used in those works. Importantly, our conditions are straightforward to verify and do not require model well-specification or the testability condition that is often invoked in prior works such as \cite{ghosal_etal.2000aos, wang_blei.2019jasa, zhang_yang.2024jrssb}. As a result, our theoretical results are applicable to a broader range of settings, including variational Bayes under model misspecification \citep{kleijn_vaart.2012ejs, wang_blei.2019nips}, and generalized Bayesian inference \citep{bissiri_etal.2016jrssb, miller.2021jmlr, knoblauch_etal.2022jmlr}.

Before presenting the main theoretical results, we provide some intuition underlying the Bernstein-von Mises (BvM) theorem for our proposed $\mvb$ estimator. Let 
\begin{equation}	\label{eq:local_max}
	\hat{\vect{\theta}}_j = \argmax{\vect{\theta}\in\Theta}\sum_{i\in\mcH_j}\ell(\vect{\theta},\vect{X}_i) =:  \argmax{\vect{\theta}\in\Theta}\mcL(\vect{\theta},\mbX_j)
\end{equation}
denote the maximizer of the local log-likelihood function. We first establish a BvM-type result for the KL projection of the $m$-powered local likelihood onto the Gaussian mean-field variational family. 

\begin{theorem}	\label{thm:mpower_BvM}
	Under Conditions \ref{cond:thick_prior} to \ref{cond:L_smooth}, let $Q^*(\hat{\vect{\theta}}_j)=\mcN(\hat{\vect{\theta}}_j,\frac{1}{mn}\diag(\vect{H})^{-1})$, and define
	\begin{equation*}
		\hat{Q}_{j}=\argmin{Q\in\mcQ^p}\kl(Q\| \tilde{P^m}(\vect{\theta}|\mbX_j)).
	\end{equation*}
	Then for every $\gamma>0$, there exists a constant $C>0$ such that
	\begin{equation}	\label{eq:mpower_BvM}
		\kl(\hat{Q}_{j}\|Q^*(\hat{\vect{\theta}}_j))\leq C\frac{m(\log n)^{3/2}}{\sqrt{n}},
	\end{equation}
	holds with probability at least $1-O(n^{-\gamma})$.
\end{theorem}

Theorem \ref{thm:mpower_BvM} provides a non-asymptotic characterization of the variational posterior corresponding to the $m$-powered local likelihood. The approximation error is bounded by $m(\log n)^{3/2}/\sqrt{n}$, which tends to zero as long as $m=o(\sqrt{n})$. This condition reflects the statistical cost of applying the $m$-power transformation. Our theoretical analysis leverages exponential-type concentration inequalities, resulting in a complementary probability of $O(n^{-\gamma})$ that decays exponentially. This represents a refinement over prior work such as \cite{han_yang.2019arXiv, zhang_yang.2024jrssb}, and enables us to rigorously handle a diverging number of subsets $m$. Theorem \ref{thm:mpower_BvM} implies that each local variational posterior converges to a Gaussian distribution $\mathcal{N}(\hat{\vect{\theta}}_j, \frac{1}{mn} \diag(\vect{H})^{-1})$, all of which share a common covariance structure. Leveraging this, we establish the following proposition concerning the $\mvb$ estimator applied to these local limiting distributions.

\begin{proposition}	\label{prop:normal_M3VB}
	Let $Q^*(\hat{\vect{\theta}}_j)\sim\mcN(\hat{\vect{\theta}}_j,\vect{\Omega}^{-1})$, and define
	\begin{equation*}
		\bar{F}_{\vect{\theta}} = \argmin{F_{\vect{\theta}}\in\mcQ}\max_{G_{\vect{\theta}}\in\mcQ}\med{1\leq j\leq m}\Big\{\kl(F_{\vect{\theta}}\|Q^*(\hat{\vect{\theta}}_j)) - \kl(G_{\vect{\theta}}\|Q^*(\hat{\vect{\theta}}_j))\Big\}.
	\end{equation*}
	\begin{itemize}
		\item If $\mcQ$ is the class of all probability distributions, then $\bar{F}_{\vect{\theta}}\sim\mcN(\vect{\mu}_{\bar{F}_{\vect{\theta}}} ,\vect{\Omega}^{-1})$, where
			\begin{equation}	\label{eq:normal_med}
				\vect{\mu}_{\bar{F}_{\vect{\theta}}} = \argmin{\vect{\theta}_f}\max_{\vect{\theta}_g}\med{1\leq j\leq m}\Big\{(\vect{\theta}_f-\hat{\vect{\theta}}_j)^{\tp}\vect{\Omega}(\vect{\theta}_f-\hat{\vect{\theta}}_j) - (\vect{\theta}_g-\hat{\vect{\theta}}_j)^{\tp}\vect{\Omega}(\vect{\theta}_g-\hat{\vect{\theta}}_j)\Big\}.
			\end{equation}
		\item If $\mcQ$ is restricted to the mean-field family, then $\bar{F}_{\vect{\theta}}\sim\mcN(\vect{\mu}_{\bar{F}_{\vect{\theta}}} ,\diag(\vect{\Omega})^{-1})$, where $\vect{\mu}_{\bar{F}_{\vect{\theta}}}$ is as defined in \eqref{eq:normal_med}.
	\end{itemize}
\end{proposition}

From the above proposition, we observe that when all local distributions are Gaussian with a shared covariance information, the $\mvb$ estimator preserves the covariance information. Its mean explicitly characterized by the min-max optimization problem in \eqref{eq:normal_med}. Furthermore, it is straightforward to verify that \eqref{eq:normal_med} is asymptotically equivalent to the empirical min-max formulation:
\begin{equation}	\label{eq:emp_minmax_med}
	\hat{\vect{\theta}} =\argmin{\vect{\theta}_f}\max_{\vect{\theta}_g}\med{1\leq j\leq m}\Big\{ \mcL(\vect{\theta}_g,\mbX_j) - \mcL(\vect{\theta}_f,\mbX_j)\Big\},
\end{equation}
by taking second-order Taylor expansion of the local log-likelihood functions $\mcL(\vect{\theta},\mbX_j)$ centered at $\hat{\vect{\theta}}_j$. The formulation in \eqref{eq:emp_minmax_med} aligns with the classical min-max estimators for empirical loss, as developed in \cite{lecue_lerasle.2019spa, lecue_lerasle.2020aos}.

For notational convenience, we define the quantile function of the loss difference as:
\begin{equation}	\label{eq:mcF_tau}
	\mcF_{\tau}(\vect{\theta}_f,\vect{\theta}_g)=\quan_{\tau}\Big\{(\vect{\theta}_f-\hat{\vect{\theta}}_j)^{\tp}\vect{H}(\vect{\theta}_f-\hat{\vect{\theta}}_j)-(\vect{\theta}_g-\hat{\vect{\theta}}_j)^{\tp}\vect{H}(\vect{\theta}_g-\hat{\vect{\theta}}_j)\Big|j\in\mcB^c\Big\},
\end{equation}
which denotes the $\tau$-th quantile computed over the uncontaminated subset indices $j\in\mcB^c$. We are now ready to present the main theoretical results of our $\mvb$ estimator. 

\begin{theorem}	\label{thm:von_mises}
Under Conditions \ref{cond:thick_prior} to \ref{cond:L_smooth}, let $Q^*(\vect{\mu}_{\hat{F}_{\vect{\theta}}}) \sim \mcN(\vect{\mu}_{\hat{F}_{\vect{\theta}}},\frac{1}{mn}\diag(\vect{H})^{-1})$. Then, with probability at least $1-O(n^{-\gamma})$, the following inequality holds:
\begin{align*}
	\kl(\hat{F}_{\vect{\theta}}\| Q^*(\vect{\mu}_{\hat{F}_{\vect{\theta}}})) \leq& C_1\frac{m(\log n)^{3/2}}{\sqrt{n}}+\frac{mn}{2}\min_{\vect{\theta}_f}\max_{\vect{\theta}_g}\mcF_{\tau}(\vect{\theta}_f,\vect{\theta}_g)\stepcounter{equation}\tag{\theequation}\label{eq:Fhat_KL_bound}\\
	& -\frac{mn}{2}\min_{\vect{\theta}_f}\max_{\vect{\theta}_g}\mcF_{1-\tau}(\vect{\theta}_f,\vect{\theta}_g),
\end{align*}
where $\tau = (1-2\alpha_n)/(2-2\alpha_n)$. Moreover, the mean $\vect{\mu}_{\hat{F}_{\vect{\theta}}}$ satisfies
\begin{align*}
	&\max_{\vect{\theta}_g} \mcF_{1-\tau}(\vect{\mu}_{\hat{F}_{\vect{\theta}}},\vect{\theta}_g) \leq \min_{\vect{\theta}_f}\max_{\vect{\theta}_g}\mcF_{\tau}(\vect{\theta}_f,\vect{\theta}_g) +C_2\frac{(\log n)^{3/2}}{n^{3/2}}. \stepcounter{equation}\tag{\theequation}\label{eq:muFhat_bound}
\end{align*}
\end{theorem}

The first inequality in Theorem \ref{thm:von_mises} provides an upper bound on the KL divergence between $\hat{F}_{\vect{\theta}}$ and the Gaussian distribution centered at its mean $\vect{\mu}_{\hat{F}_{\vect{\theta}}}$. To ensure this divergence remains of order $o(1)$, it is required that $m=o(\sqrt{n})$. The last two terms in \eqref{eq:Fhat_KL_bound} quantify the discrepancy between the $\tau$-th and $(1-\tau)$-th quantiles of the empirical loss difference $\{\mcL(\vect{\theta}_f,\mbX_j)-\mcL(\vect{\theta}_g,\mbX_j)\}_{j=1}^m$. According to Theorem \ref{thm:theta_ftau} below, we have $\min_{\vect{\theta}_f}\max_{\vect{\theta}_g}\mcF_{\tau}(\vect{\theta}_f,\vect{\theta}_g)= O_{\mbP}\big(1/(mn)\big)$, under the condition $\alpha_n=O(1/\sqrt{m})$. Consequently, in \eqref{eq:Fhat_KL_bound}, the KL divergence between $\hat{F}_{\vect{\theta}}$ and the reference distribution $Q^*(\vect{\mu}_{\hat{F}_{\vect{\theta}}})$ is of order $O(1)$. Noting that $Q^*(\vect{\mu}_{\hat{F}_{\vect{\theta}}})$ has variance of order $O(1/(mn))$, a KL divergence of order $O(1)$  is sufficient to establish concentration of $\hat{F}_{\vect{\theta}}$ around its mean (see Lemma 5 in the Appendix). Equation \eqref{eq:muFhat_bound} provides an indirect control of the mean of $\hat{F}_{\vect{\theta}}$. A more explicit statistical rate for $\vect{\mu}_{\hat{F}_{\vect{\theta}}}$ will be established in the next section. The expression for $\tau = \frac{1 - 2\alpha_n}{2 - 2\alpha_n}$ arises from the observation that, for any sequence $\{x_j\}_{j=1}^m$ with a corrupted index set $\mcB$, the following inequality holds:
\begin{equation*}
	\quan_{1-\tau}(x_i|j\notin\mcB)\leq \med{1\leq j\leq m}(x_i)\leq \quan_{\tau}(x_i|j\notin\mcB),
\end{equation*} 
implying that the global median can be controlled via the quantiles of the uncontaminated subsets. We conclude this section with a corollary for the case without contamination, i.e., when $\alpha_n=0$.

\begin{corollary}
	Under the same conditions as in Theorem \ref{thm:von_mises}, and assuming $\alpha_n=0$, the following inequality holds:
\begin{align*}
	\kl(\hat{F}_{\vect{\theta}}\| Q^*(\vect{\mu}_{\hat{F}_{\vect{\theta}}})) \leq& C_1\frac{m(\log n)^{3/2}}{\sqrt{n}},
\end{align*}
where $\vect{\mu}_{\hat{F}_{\vect{\theta}}}$ satisfies
\begin{align*}
	&\max_{\vect{\theta}_g} \mcF_{1/2}(\vect{\mu}_{\hat{F}_{\vect{\theta}}},\vect{\theta}_g) \leq \min_{\vect{\theta}_f}\max_{\vect{\theta}_g}\mcF_{1/2}(\vect{\theta}_f,\vect{\theta}_g)+C_2\frac{(\log n)^{3/2}}{n^{3/2}}.
\end{align*}
\end{corollary}

This corollary indicates that, in the absence of data corruption, the KL divergence between the aggregated variational posterior $\hat{F}_{\vect{\theta}}$ and the Gaussian distribution $Q^*(\vect{\mu}_{\hat{F}_{\vect{\theta}}})$ is of order $o(1)$, provided that $m=o(\sqrt{n})$. 


\subsection{Statistical Rate of the Mean of the $\mvb$ Estimator}	\label{sec:mu_rate}

In this section, we aim to derive a precise statistical convergence rate for the mean of the $\mvb$ estimator, denoted as $\vect{\mu}_{\hat{F}_{\vect{\theta}}}$, which is governed by the bound in \eqref{eq:muFhat_bound}. As previously discussed, $\vect{\mu}_{\hat{F}_{\vect{\theta}}}$ closely approximates the solution to the min-max empirical loss problem in \eqref{eq:emp_minmax_med}. According to existing theoretical results on min-max median estimator \citep{lecue_lerasle.2019spa, lecue_lerasle.2020aos}, the estimator $\hat{\vect{\theta}}$ achieves a convergence rate $O_{\mbP}(\sqrt{\alpha_n/n})$. In this section, we endeavor to establish an improved convergence rate for $\vect{\mu}_{\hat{F}_{\vect{\theta}}}$. 

To facilitate a clear exposition of our theoretical contributions, we first consider the simplified setting in which no data corruption is present; that is, $\alpha_n=0$ and hence $\tau=1/2$. In this case, the function $\mcF_{1/2}(\vect{\theta}_f,\vect{\theta}_g)$ simplifies to:
\begin{align*}
	\mcF_{0.5}(\vect{\theta}_f,\vect{\theta}_g) =& (\vect{\theta}_f-\vect{\theta}^*)^{\tp}\vect{H}(\vect{\theta}_f-\vect{\theta}^*)-(\vect{\theta}_g-\vect{\theta}^*)^{\tp}\vect{H}(\vect{\theta}_g-\vect{\theta}^*)\\
		&-\med{1\leq j\leq m}\Big\{2(\vect{\theta}_f-\vect{\theta}_g)^{\tp}\vect{H}(\hat{\vect{\theta}}_j - \vect{\theta}^*)\Big\}.
\end{align*}
Using the Bahadur representation of the local maximizer $\hat{\vect{\theta}}_j$, we have:
\begin{equation*}
	\hat{\vect{\theta}}_j - \vect{\theta}^* = \vect{H}^{-1}\frac{1}{n}\sum_{i\in\mcH_j}\nabla\ell(\vect{X}_i,\vect{\theta}^*) + O_{\mbP}\Big(\frac{\log n}{n}\Big),
\end{equation*}
where the leading term is the empirical mean of i.i.d. zero-mean random vectors. By the Berry-Esseen theorem, the collection of random variables $\{2(\vect{\theta}_f-\vect{\theta}_g)^{\tp}\vect{H}(\hat{\vect{\theta}}_j - \vect{\theta}^*)\}_{j=1}^m$ is approximately symmetric and independent. Consequently, applying standard arguments from the theory of median-of-means estimator \citep{minsker.2019ejs}, we obtain that the median term achieves a sharper convergence rate of $O_{\mbP}(|\vect{\theta}_f-\vect{\theta}_g|_2/\sqrt{mn})$. This observation enables us to establish an improved statistical rate for the min-max estimator, outperforming the rate obtained in \cite{lecue_lerasle.2019spa, lecue_lerasle.2020aos}.

Before analyzing the statistical rate of the mean $\vect{\mu}_{\hat{F}_{\vect{\theta}}}$, we introduce and study an intermediate parameter defined by 
\begin{align*}
	\hat{\vect{\theta}}_{f,\tau} =&\argmin{\vect{\theta}_f}\max_{\vect{\theta}_g} \mcF_{\tau}(\vect{\theta}_f,\vect{\theta}_g).\stepcounter{equation}\tag{\theequation}\label{eq:hattheta_ftau_def}
\end{align*}
We now establish the statistical properties of $\hat{\vect{\theta}}_{f,\tau}$ through the following theorem.

\begin{theorem}	\label{thm:theta_ftau}
	Suppose Conditions \ref{cond:thick_prior} to \ref{cond:L_smooth} hold. Let $\tau\in[\delta,1-\delta]$ for some fixed constant $\delta>0$. Then the estimator $\hat{\vect{\theta}}_{f,\tau}$ defined in \eqref{eq:hattheta_ftau_def} satisfies
	\begin{equation*}
		|\hat{\vect{\theta}}_{f,\tau} - \vect{\theta}^*|_2\leq \frac{C_1}{\sqrt{n}}\Big(\Big|\tau-\frac{1}{2}\Big|+\sqrt{\frac{\log n}{m}}+\frac{\log n}{\sqrt{n}}\Big).
	\end{equation*}
	and
	\begin{equation}	\label{eq:Ftau_thetafg}
		\max_{\vect{\theta}_g}\mcF_{\tau}(\hat{\vect{\theta}}_{f,\tau},\vect{\theta}_g)\leq \frac{C_2}{n}\Big(\Big|\tau-\frac{1}{2}\Big|^2+\frac{\log n}{m}+\frac{(\log n)^2}{n}\Big).
	\end{equation}
\end{theorem}

Theorem \ref{thm:theta_ftau} establishes the statistical convergence rate of the min-max quantile estimator $\hat{\vect{\theta}}_{f,\tau}$. In general, this estimator achieves a convergence rate of order $O(1/\sqrt{n})$ for any $\tau\in[\delta,1-\delta]$, which is consistent with the results reported in \cite{lecue_lerasle.2019spa, lecue_lerasle.2020aos}. Notably, when $m=O(n)$ and $|\tau-1/2|=O(1/\sqrt{m})$, the estimator $\hat{\vect{\theta}}_{f,\tau}$ attains a nearly optimal rate of $O(\sqrt{\log n/(mn)})$, which is not covered by existing theoretical analyses. Equation \eqref{eq:Ftau_thetafg} provides an upper bound for the quantity $\max_{\vect{\theta}_g}\mcF_{\tau}(\hat{\vect{\theta}}_{f,\tau},\vect{\theta}_g)$, which is crucial for deriving the statistical rate of $\vect{\mu}_{\hat{F}_{\vect{\theta}}}$ via the control provided in equation \eqref{eq:muFhat_bound}. The result is formalized in the following theorem:

\begin{theorem}	\label{thm:mu_hatF_rate}
	Suppose Conditions \ref{cond:thick_prior} to \ref{cond:L_smooth} hold. Then for any $\alpha_n\in[0,1/2-\delta]$ where $\delta>0$ is fixed, the mean $\vect{\mu}_{\hat{F}_{\vect{\theta}}}$ defined via equation \eqref{eq:muFhat_bound} satisfies the bound:
	\begin{equation}	\label{eq:mu_hatF_rate}
		|\vect{\mu}_{\hat{F}_{\vect{\theta}}} - \vect{\theta}^*|_2\leq C\Big(\frac{\alpha_n}{\sqrt{n}}+\sqrt{\frac{\log n}{mn}}+\frac{(\log n)^{3/4}}{n^{3/4}}\Big).
	\end{equation}
\end{theorem}

Compared to the min-max quantile estimator $\hat{\vect{\theta}}_{f,\tau}$ in equation \eqref{eq:hattheta_ftau_def}, the convergence rate for $\vect{\mu}_{\hat{F}_{\vect{\theta}}}$ is slightly worse in the third term by a factor of $n^{1/4}$. This discrepancy arises because the term $\max_{\vect{\theta}_g} \mcF_{1-\tau}(\vect{\mu}_{\hat{F}_{\vect{\theta}}},\vect{\theta}_g)$ in \eqref{eq:muFhat_bound} incurs an additional approximation error of order $O((\log n)^{3/2}/n^{3/2})$. Such $O(n^{-3/4})$ error term also appear in prior literature, for example, see Corollary 1 of \cite{han_yang.2019arXiv}. When $m=O(\sqrt{n})$, the estimator $\vect{\mu}_{\hat{F}_{\vect{\theta}}}$ achieves a nearly optimal statistical rate. Furthermore, the rate we obtain improves upon the $O(1/\sqrt{n})$ rate established in \cite{minsker_etal.2017jmlr, padilla_etal.2025arXiv}, which rely on aggregating local (variational) posteriors via geometric medians in the space of probability distributions.

\begin{remark}  \label{rem:mvb_mean}
	Suppose we consider the estimator $\hat{Q}_{\vect{\theta}}$ defined via the median-based minimization in equation \eqref{eq:naive_KLVB_median}. One can similarly establish a Bernstein-von Mises theorem for $\hat{Q}_{\vect{\theta}}$, under which its mean $\vect{\mu}_{\hat{Q}_{\vect{\theta}}}$ is expected to be close to the minimizer of the following objective:
	\begin{align*}
		&\min_{\vect{\theta}}\Big[\med{ 1\leq j\leq m}\Big\{(\vect{\theta}-\hat{\vect{\theta}}_j)^{\tp}\vect{H}(\vect{\theta}-\hat{\vect{\theta}}_j)\Big\}\Big]\\
		=&\min_{\vect{\theta}}\Big[(\vect{\theta}-\vect{\theta}^*)^{\tp}\vect{H}(\vect{\theta}-\vect{\theta}^*)+\med{1\leq j\leq m}\Big\{2(\vect{\theta}-\vect{\theta}^*)^{\tp}\vect{H}(\vect{\theta}^*-\hat{\vect{\theta}}_j)+(\vect{\theta}^*-\hat{\vect{\theta}}_j)^{\tp}\vect{H}(\vect{\theta}^*-\hat{\vect{\theta}}_j)\Big\}\Big],
	\end{align*}
	in the absence of contamination. However, the bracketed terms inside the median are no longer jointly asymptotically symmetric. As a result, applying the median directly introduces bias in the estimation. In contrast, our proposed min-max formulation eliminates the last term and retains only the nearly symmetric term $(\vect{\theta}-\vect{\theta}^*)\vect{H}(\vect{\theta}^*-\hat{\vect{\theta}}_j)$ within the median. This structural simplification substantially facilitates the theoretical analysis. While a more refined analysis may be conducted to quantify the bias and derive a precise statistical rate for $\vect{\mu}_{\hat{Q}_{\vect{\theta}}}$, such investigation is beyond the scope of the current paper. We leave this as an interesting direction for future research.
\end{remark}


\section{$\mvb$ with Local Latent Variables}	\label{sec:med_elbo}

\paragraph{Variational Bayes with Local Latent Variables} In certain settings, such as Gaussian mixture models, the introduction of local latent variables can substantially simplify the evaluation of the likelihood function. In this section, we extend the min-max median variational Bayes ($\mvb$) framework to accommodate models with local latent variables. Consider a dataset $\mbX=\{\vect{X}_1,...,\vect{X}_N\}$. For each observed covariate $\vect{X}_i$, there exists an associated unobserved local latent variable $S_i\in\mcS$. The joint distribution $P(\mbX,\mbS,\vect{\theta})$ has density function
\begin{equation*}
	p(\mbX,\mbS,\vect{\theta}) = \pi(\vect{\theta})p(\mbX,\mbS|\vect{\theta}) = \pi(\vect{\theta})\prod_{i=1}^N\big\{p(\vect{X}_i|\vect{\theta},S_i)p(S_i|\vect{\theta})\big\}.
\end{equation*}
The marginal likelihood of the observed data $\mbX$ is then defined by integrating out the local latent variables:
\begin{equation*}
	p(\mbX|\vect{\theta}) = \int_{\mcS^n}p(\mbX|\vect{\theta},\mbS)p(\mbS|\vect{\theta})\diff \mbS.
\end{equation*}
In latent variable model of this form, our objective is to approximate the marginal posterior distribution $p(\vect{\theta}|\mbX)$ of the global parameter $\vect{\theta}$ based on the joint posterior distribution $P(\vect{\theta},\mbS|\mbX)$ over the global parameter $\vect{\theta}$ and local latent variables $\mbS$ has density function
\begin{align*}
	p(\vect{\theta},\mbS|\mbX) = \frac{p(\mbX|\mbS,\vect{\theta})p(\mbS|\vect{\theta})\pi(\vect{\theta})}{p(\mbX)}.
\end{align*}
Noticing that the normalizing constant $p(\mbX)$ is typically intractable to compute in practice. The variational Bayes framework addresses the issue by approximating the joint posterior through optimization. Specifically, VB solves: 
\begin{align*}
	\hat{Q}_{\vect{\theta}} =& \argmin{Q_{\vect{\theta}}\in\mcQ^p}\min_{Q_{\mbS}\in\mcQ^N}\kl(Q_{\vect{\theta}}\otimes Q_{\mbS}\| P(\mbS,\vect{\theta}|\mbX))\stepcounter{equation}\tag{\theequation}\label{eq:elbo_ll_max}\\
		=&\argmax{Q_{\vect{\theta}}\in\mcQ^p}\max_{Q_{\mbS}\in\mcQ^N}\elbo\big(Q_{\vect{\theta}}\otimes Q_{\mbS}\| P(\mbS,\vect{\theta}, \mbX)\big).
\end{align*}
Compared to the VB setting without local latent variables as described in equation \eqref{eq:elbo_max}, we need to also approximate the distribution of the local latent variables $\mbS$. 

\paragraph{Inconsistency of One-Stage $\mvb$} Following the methodology presented in Section \ref{sec:mean_dist}, we divide the dataset $\mbX$ into $m$ subsets $\mbX_j=\{\vect{X}_i|i\in\mcH_j\}$, where $\mcH_j$, for $1\leq j\leq m$, denotes the corresponding index set for the $j$-th subset. Each subset $\mbX_j$ is associated with a corresponding set of local latent variables denoted by $\mbS_j$. To construct a robust aggregator in this context, we naturally extend the formulation of equation \eqref{eq:minmax_median} to the setting with local latent variables as follows:
\begin{align*}
	\hat{F}'_{\vect{\theta}}=& \argmin{F_{\vect{\theta}}\in\mcQ^{p}}\min_{F_{\mbS}\in\mcQ^{mn}}\max_{G\in\mcQ^{p+mn}}\Big[\med{1\leq j\leq m}\Big\{\kl(F_{\vect{\theta}}\otimes F_{_{\mbS_j}}\| \tilde{P^m}(\vect{\theta},\mbS_j|\mbX_j))\\
	& \qquad\qquad- \kl(G_{\vect{\theta}}\otimes G_{_{\mbS_j}}\| \tilde{P^m}(\vect{\theta},\mbS_j|\mbX_j))\Big\}\Big]\\
	=& \argmin{F_{\vect{\theta}}\in\mcQ^{p}}\min_{F_{\mbS}\in\mcQ^{mn}}\max_{G\in\mcQ^{p+mn}}\Big[\med{1\leq j\leq m}\Big\{\elbo(G_{\vect{\theta}}\otimes G_{_{\mbS_j}}\| \tilde{P^m}(\vect{\theta},\mbS_j,\mbX_j))\\
	&  \qquad\qquad-\elbo(F_{\vect{\theta}}\otimes F_{_{\mbS_j}}\| \tilde{P^m}(\vect{\theta},\mbS_j,\mbX_j))\Big\}\Big],\stepcounter{equation}\tag{\theequation}\label{eq:minmax_median_naive}
\end{align*}
where $\tilde{P^m}(\vect{\theta},\mbS_j,\mbX_j)$ denotes the $m$-powered likelihood, with density function
\begin{equation*}
	\tilde{p^m}(\vect{\theta},\mbS_j,\mbX_j) \propto \pi(\vect{\theta})p^m(\mbX|\mbS,\vect{\theta})p^m(\mbS|\vect{\theta}).
\end{equation*}

An intriguing phenomenon arises when incorporating local latent variables. By slightly adapting the theoretical framework developed in Section \ref{sec:theory_M3VB}, we can show that the solution $\hat{F}'(\vect{\theta})$ to \eqref{eq:minmax_median_naive} has a mean $\vect{\mu}_{\hat{F}'_{\vect{\theta}}}$ that satisfies
\begin{equation}	\label{eq:thetastar_m}
	\vect{\mu}_{\hat{F}'_{\vect{\theta}}}\rightarrow\vect{\theta}^*_m = \argmax{\vect{\theta}}\mbE[\log(\tilde{p^m}(\vect{X}|\vect{\theta}))],
\end{equation}
where
\begin{equation*}
	\tilde{p^m}(\vect{X}|\vect{\theta})\propto \int_{\mcS}p^m(\vect{X},S|\vect{\theta})\diff S = \int_{\mcS}p^m(\vect{X}|S, \vect{\theta})p^m(S|\vect{\theta})\diff S .
\end{equation*}
Surprisingly, in the presence of local latent variables $S$, this expression is generally \emph{not} proportional to the $m$-th power of the marginal likelihood, i.e., 
\begin{equation*}
	\big\{p(\vect{X}|\vect{\theta})\big\}^m = \Big\{\int_{\mcS}p(\vect{X},S|\vect{\theta})\diff S\Big\}^m.
\end{equation*}
This discrepancy leads to the important conclusion that
\begin{equation*}
	\vect{\theta}^*_m\neq \vect{\theta}^* = \argmax{\vect{\theta}}\mbE[\log(p(\vect{X}|\vect{\theta}))].
\end{equation*}
In summary, the mean of $\hat{F}'(\vect{\theta})$ is, in general, not consistent when the KL divergence is minimized with respect to the $m$-powered posterior distribution. In particular, the population-level minimizer $\vect{\theta}_m^*$ defined in \eqref{eq:thetastar_m} is not invariant with respect to the choice of the power $m$. To illustrate this phenomenon, we provide a concrete example demonstrating the discrepancy:

\begin{example}	\label{exp:inco_pm}
	Let $\Theta=\{0,1\},\mcX=\{0,1\},\mcS=\{0,1\}$, and assume that the true parameter is $\theta^*=0$. Define the data-generating process via the following conditional probabilities: 
	\begin{align*}
		&\mbP(S=0|\theta=0) = \frac{3}{8},\quad \mbP(S=0|\theta=1) = \frac{1}{2},\\
		&\mbP(X=0|S=0,\theta=0) = \frac{5}{6},\quad \mbP(X=0|S=1,\theta=0) = \frac{2}{5},\\
		&\mbP(X=0|S=0,\theta=1) = \frac{11}{16},\quad \mbP(X=0|S=1,\theta=1) = \frac{1}{4}.
	\end{align*}
	Under this setup, we have the following result:
	\begin{align*}
		\argmax{\theta}\mbE_{\theta^*}[\log p(X|\theta)]=\theta^* \neq \argmax{\theta}\lim_{m\rightarrow\infty}\mbE_{\theta^*}[\log \tilde{p^m}(X|\theta)].
	\end{align*}
\end{example}

\paragraph{Two-Stage $\mvb$} Based on the preceding analysis, it becomes apparent that applying the $m$-th power to the likelihood function is generally inappropriate when local latent variables are present. To address this, we propose an alternative approach: bypassing the $m$-power transformation and instead directly solving the following min-max optimization problem:
\begin{align*}
	\tilde{F}_{\vect{\theta}}=& \argmin{F_{\vect{\theta}}\in\mcQ^{p}}\min_{F_{\mbS}\in\mcQ^{mn}}\max_{G\in\mcQ^{p+mn}}\Big[\med{1\leq j\leq m}\Big\{\kl(F_{\vect{\theta}}\otimes F_{_{\mbS_j}}\| P(\vect{\theta},\mbS_j|\mbX_j))\\
	&\qquad\qquad - \kl(G_{\vect{\theta}}\otimes G_{_{\mbS_j}}\| P(\vect{\theta},\mbS_j|\mbX_j))\Big\}\Big]\\
	=& \argmin{F_{\vect{\theta}}\in\mcQ^{p}}\min_{F_{\mbS}\in\mcQ^{mn}}\max_{G\in\mcQ^{p+mn}}\Big[\med{1\leq j\leq m}\Big\{\elbo(G_{\vect{\theta}}\otimes G_{_{\mbS_j}}\| P(\vect{\theta},\mbS_j,\mbX_j))\\
	&\qquad\qquad-\elbo(F_{\vect{\theta}}\otimes F_{_{\mbS_j}}\| P(\vect{\theta},\mbS_j,\mbX_j))\Big\}\Big].\stepcounter{equation}\tag{\theequation}\label{eq:minmax_median_ll}
\end{align*}
As discussed in Section \ref{sec:mean_dist}, the resulting distribution $\tilde{F}_{\vect{\theta}}$ typically exhibits an asymptotic covariance on the order of $O(m/N)$, which is substantially larger than that of the full posterior. To correct for this inflation and recover statistical efficiency, we propose a rescaling procedure: we reduce the variance of $\tilde{F}_{\vect{\theta}}$ by a factor of $1/m$ to obtain the final aggregated distribution $\hat{F}_{\vect{\theta}}$. For many common distribution families-such as the Gaussian or fully factorized Gaussian-this rescaling can be achieved by simply multiplying the covariance matrix by a factor of $1/m$. For general distribution families, the density function $\tilde{f}_{\vect{\theta}}$ can be rescaled around its mean $\vect{\mu}_{\tilde{F}_{\vect{\theta}}}$ as follows:
\begin{equation}	\label{eq:ftilde_rescale}
	\hat{f}_{\vect{\theta}}(\vect{\theta}) = m^{p/2}\tilde{f}\big\{\vect{\mu}_{\tilde{F}_{\vect{\theta}}}+\sqrt{m}(\vect{\theta} - \vect{\mu}_{\tilde{F}_{\vect{\theta}}})\big\}.
\end{equation}
It is straightforward to verify that the rescaled density function $\hat{f}_{\vect{\theta}}(\vect{\theta})$ retains the same mean as the intermidiate density $\tilde{f}_{\vect{\theta}}(\vect{\theta})$, while its variance is reduced by a factor of $1/m$.


\subsection{Theory of $\mvb$ with Local Latent Variables}	\label{sec:theory_M3VB_ll}

In this section, we develop the theoretical properties of the $\mvb$ estimator in the presence of local latent variables. To this end, we introduce additional technical conditions. Let $\ell_{S}(\vect{\theta},\vect{X}) = \log p(S|\vect{\theta},\vect{X})$ denote the conditional log-density of the local latent variable given the covariates and global parameter. We define the corresponding negative Hessian matrix $\vect{H}_s = -\mbE[\nabla^2\ell_S(\vect{\theta}^*,\vect{X})]$, and the complete information matrix as $\vect{H}_c = \vect{H} +\vect{H}_s$.

\begin{condition}	\label{cond:S_exp_hess}
	Denote $p_{s}(\vect{\theta},\vect{X}) := p(S=s|\vect{X},\vect{\theta})$, there exist constants $\eta_4,M_4>0$ such that
	\begin{equation*}
		\max_{1\leq l\leq p}\sup_{\vect{v}\in\mbS^{p-1}}\sup_{\vect{\theta}\in\Theta}\mbE\Big[\exp\Big\{\eta_4\int_{\mcS}p_s(\vect{\theta},\vect{X})\vect{v}^{\tp}\nabla^2\ell_{s}(\vect{\theta},\vect{X})\vect{v}\diff s\Big\}\Big]<M_4.
	\end{equation*}
\end{condition}

\begin{condition}	\label{cond:SL_smooth}
	The function $\ell_S(\vect{\theta},\vect{X})$ is three times continuously differentiable. Moreover, there exists a constant $L_2>0$ such that
	\begin{align*}
		\Big|\frac{\partial^3}{\partial\theta_i\partial\theta_j\partial\theta_k}\ell_S(\vect{\theta},\vect{X})\Big|\leq L_2,\quad 1\leq i,j,k\leq p.
	\end{align*}
\end{condition}

Condition \ref{cond:S_exp_hess} imposes a sub-exponential tail bound on the Hessian of $\ell_S(\vect{\theta},\vect{X})$. In \cite{han_yang.2019arXiv, zhang_yang.2024jrssb}, the authors assume only moment conditions on the Hessian, which are weaker than ours. Although Condition \ref{cond:S_exp_hess} may appear intricate due to its integral over the latent space $\mcS$, in practical scenarios such as Gaussian mixture models where $\mcS$ is finite, the condition reduces to verifying sub-exponential behavior for each discrete state $s\in\mcS$ individually. Condition \ref{cond:SL_smooth} requires third-order differentiability of $\ell_S(\vect{\theta}, \vect{X})$, which is a slightly stronger regularity assumption than those made in \cite{han_yang.2019arXiv, zhang_yang.2024jrssb}. Similar to Condition \ref{cond:L_smooth}, this assumption is primarily imposed to facilitate technical derivations in our theoretical analysis. Given these conditions, we now present the Bernstein-von Mises theorem for the local variational posterior.

\begin{theorem}	\label{thm:BvM_ll}
	Under Conditions \ref{cond:thick_prior} to \ref{cond:SL_smooth}, define $Q^*(\hat{\vect{\theta}}_j)=\mcN(\hat{\vect{\theta}}_j,\frac{1}{n}\diag(\vect{H}_c)^{-1})$, and consider the local variational approximation
	\begin{equation*}
		\hat{Q}_{j,\vect{\theta}}=\argmin{Q_{\vect{\theta}}\in\mcQ^p}\min_{Q_{\mbS_j}\in\mcQ^n}\kl(Q_{\vect{\theta}}\otimes Q_{\mbS_j}\| P(\vect{\theta},\mbS_j|\mbX_j)).
	\end{equation*}
	Then, for any $\gamma>0$, there exists a constant $C>0$ such that
	\begin{equation*}
		\kl(\hat{Q}_{j,\vect{\theta}}\|Q^*(\hat{\vect{\theta}}_j))\leq C\frac{(\log n)^{3/2}}{\sqrt{n}},
	\end{equation*}
	with probability at least $1-O(n^{-\gamma})$.
\end{theorem}

Similarly to Theorem \ref{thm:mpower_BvM}, this result refines existing Bernstein-von Mises theorem by strengthening the tail bound to an exponentailly decaying probability. In the following, we turn to the theoretical analysis of $\tilde{F}_{\vect{\theta}}$, as defined in \eqref{eq:minmax_median_ll}.

\begin{theorem}	\label{thm:von_mises_ll}
Suppose Conditions \ref{cond:thick_prior} to \ref{cond:SL_smooth} hold. Let $Q^*(\vect{\mu}_{\tilde{F}_{\vect{\theta}}})\sim \mcN(\vect{\mu}_{\tilde{F}_{\vect{\theta}}},\frac{1}{n}\diag(\vect{H}_c)^{-1})$, then with probability at least $1-O(n^{-\gamma})$, the following bound holds:
\begin{align*}
	\kl(\tilde{F}_{\vect{\theta}}\| Q^*(\vect{\mu}_{\tilde{F}_{\vect{\theta}}}))\leq &C_1\frac{(\log n)^{3/2}}{\sqrt{n}}+\frac{n}{2}\min_{\vect{\theta}_f}\max_{\vect{\theta}_g}\mcF_{\tau}(\vect{\theta}_f,\vect{\theta}_g)\stepcounter{equation}\tag{\theequation}\label{eq:Fhat_KL_boundll}\\
	& -\frac{n}{2}\min_{\vect{\theta}_f}\max_{\vect{\theta}_g}\mcF_{1-\tau}(\vect{\theta}_f,\vect{\theta}_g),
\end{align*}
where $\tau = (1-2\alpha_n)/(2-2\alpha_n)$ and the mean $\vect{\mu}_{\tilde{F}_{\vect{\theta}}}$ satisfies
\begin{align*}
	|\vect{\mu}_{\tilde{F}_{\vect{\theta}}} - \vect{\theta}^*|_2\leq C\Big(\frac{\alpha_n}{\sqrt{n}}+\sqrt{\frac{\log n}{mn}}+\frac{(\log n)^{3/4}}{n^{3/4}}\Big).
\end{align*}
\end{theorem}

The result closely parallels Theorem \ref{thm:von_mises}, but with all multiplicative factors involving $m$ removed. The key distinction lies in the covariance structure: the covariance of $\tilde{F}_{\vect{\theta}}$ now converges to $\frac{1}{n}\diag(\vect{H}_c)^{-1}$, where the complete Hessian matrix $\vect{H}_c$ replaces $\vect{H}$ in Theorem \ref{thm:von_mises}. Interestingly, the definition of the quantile-based function $\mcF_{\tau}(\vect{\theta}_f,\vect{\theta}_g)$-given in equation \eqref{eq:mcF_tau}-remains unchanged and still relies on the original Hessian matrix $\vect{H}$. Using this result, we now establish theoretical guarantees for the rescaled estimator $\hat{F}_{\vect{\theta}}$. 

\begin{corollary}	\label{cor:BvM_ll}
	Under the same conditions as in Theorem \ref{thm:von_mises_ll}, denote $Q^*(\vect{\mu}_{\hat{F}_{\vect{\theta}}})\sim \mcN(\vect{\mu}_{\hat{F}_{\vect{\theta}}},\frac{1}{mn}\diag(\vect{H}_c)^{-1})$. Then with probability at least $1-O(n^{-\gamma})$, the rescaled distribution $\hat{F}_{\vect{\theta}}$ satisfies
	\begin{align*}
		\kl(\hat{F}_{\vect{\theta}}\| Q^*(\vect{\mu}_{\hat{F}_{\vect{\theta}}})) \leq& C_1\frac{(\log n)^{3/2}}{\sqrt{n}}+\frac{n}{2}\min_{\vect{\theta}_f}\max_{\vect{\theta}_g}\mcF_{\tau}(\vect{\theta}_f,\vect{\theta}_g)\stepcounter{equation}\tag{\theequation}\label{eq:Fhat_KL_bound_ll}\\
	&-\frac{n}{2}\min_{\vect{\theta}_f}\max_{\vect{\theta}_g}\mcF_{1-\tau}(\vect{\theta}_f,\vect{\theta}_g),
	\end{align*}
	where the mean $\vect{\mu}_{\hat{F}_{\vect{\theta}}}$ satisfies
	\begin{equation*}
		|\vect{\mu}_{\hat{F}_{\vect{\theta}}} - \vect{\theta}^*|_2\leq C_2\Big(\frac{\alpha_n}{\sqrt{n}}+\sqrt{\frac{\log n}{mn}}+\frac{(\log n)^{3/4}}{n^{3/4}}\Big).
	\end{equation*}
\end{corollary}

It is noteworthy that the KL divergence bound in \eqref{eq:Fhat_KL_bound_ll} is not inflated by an additional factor of $m$, despite the rescaling of $\tilde{f}_{\vect{\theta}}$. This indicates that the rescaled distribution $\hat{F}_{\vect{\theta}}$ retains favorable approximation accuracy and offers superior statistical efficiency compared to directly minimizing the KL divergence from the $m$-powered local likelihood functions as in \eqref{eq:minmax_median}. This corollary also suggests that, even in the absence of local latent variables, the aggregate-and-rescale approach exhibits better theoretical performance than directly aggregating the $m$-powered local likelihood functions.


\section{Algorithms and Examples}	\label{sec:alg}

In this section, we explore algorithms for solving the min-max median variational Bayes problems defined in equations \eqref{eq:minmax_median} and \eqref{eq:minmax_median_ll}. The standard variational Bayes framework typically employs coordinate ascent methods \citep{bishop.2006, wright.2015mp} to derive the variational posterior. To adapt these traditional algorithms to the min-max median framework, we need modifications proposed by \cite{lecue_lerasle.2020aos}. Recognizing the differences in our $\mvb$ method when dealing with variational Bayes models with or without local latent variables, we will discuss these two scenarios separately. Several examples are provided to illustrate the algorithms effectively.


\subsection{$\mvb$ Algorithm without Local Latent Variable}	\label{sec:alg_withoutll}

As discussed in Section \ref{sec:mean_dist}, when local latent variables are absent, it is appropriate to raise the likelihood function $P(\vect{\theta},\mbX_j)$ to the power of $m$. From a computational perspective, define
\begin{equation*}
	\tilde{p^m}(\vect{\theta},\mbX_j) = C_mp^m(\mbX_j|\vect{\theta})\pi(\vect{\theta}),
\end{equation*} 
where $C_m$ is a normalization constant. For each component in the median, the local $\elbo$ can be written explicitly as
\begin{align*}
	\elbo \big(F(\vect{\theta})\|\tilde{P^m}(\vect{\theta},\mbX_j)\big)=&\int_{\Theta} f(\vect{\theta})\log\Big\{\frac{\tilde{p^m}(\vect{\theta},\mbX_j)}{f(\vect{\theta})}\Big\}\diff\vect{\theta}\\
		=&\underbrace{\int_{\Theta} f(\vect{\theta})\log\Big\{\frac{p^m(\mbX_j|\vect{\theta})\pi(\vect{\theta})}{f(\vect{\theta})}\Big\}\diff\vect{\theta}}_{I_j} -\log(C_m),
\end{align*}
where the constant term $\log C_m$ vanishes in the min-max objective \eqref{eq:minmax_median}. Consequently, when solving the optimization problem \eqref{eq:minmax_median}, we may equivalently maximize the term $I_j$ using the unnormalized density $p^m(\mbX_j|\vect{\theta})\pi(\vect{\theta})$. 

To adapt the min-max median framework, we incorporate the methodology proposed by \cite{lecue_lerasle.2020aos}, detailed as follows. For notational convenience, define 
\begin{equation*}
	\elbo_j(F(\vect{\theta})) = \elbo(F(\vect{\theta})\|\tilde{P^m}(\vect{\theta},\mbX_j)).
\end{equation*}
At iteration $t$, suppose the current distributions $F^{(t)}(\vect{\theta})$ and $G^{(t)}(\vect{\theta})$ have density functions
\begin{equation*}
	f^{(t)}(\vect{\theta}) = \prod_{l=1}^pf_l^{(t)}(\theta_l),\quad G^{(t)}(\vect{\theta}) = \prod_{l=1}^pg_l^{(t)}(\theta_l).
\end{equation*}
We first identify the index $j_t$ corresponding to the median block, specifically,
\begin{equation*}
	\elbo_{j_t}(G^{(t)}(\vect{\theta})) - \elbo_{j_t}(F^{(t)}(\vect{\theta})) = \med{1\leq j\leq m}\Big\{ \elbo_{j}(G^{(t)}(\vect{\theta})) - \elbo_{j}(F^{(t)}(\vect{\theta})) \Big\}.
\end{equation*}

We then perform a one-step coordinate accent update within the $j_t$-th block. Specifically, for each coordinate $l$, we update the variational factor as
\begin{equation}	\label{eq:cavi}
	f_{l}^{(t+1)}(\theta_l) \propto \exp\big\{\mbE_{-l,F^{(t)}}[\log \{p^m(\mbX_j|\vect{\theta})\pi(\vect{\theta})\}]\big\},
\end{equation}
where $\mbE_{-l, F^{(t)}}$ denotes the expectation with respect to all coordinates of $\vect{\theta}$ except the $l$-th, under the current distribution $F^{(t)}$. After updating $F^{(t+1)}(\vect{\theta})$ via the product $f^{(t+1)}(\vect{\theta})=\prod_{l=1}^pf_l^{(t+1)}(\theta_l)$, we proceed to update the distribution $G$. To this end, identify the index $j'_t$ such that
\begin{equation*}
	\elbo_{j'_t}(G^{(t)}(\vect{\theta})) - \elbo_{j'_t}(F^{(t+1)}(\vect{\theta})) = \med{1\leq j\leq m}\Big\{ \elbo_{j}(G^{(t)}(\vect{\theta})) - \elbo_{j}(F^{(t+1)}(\vect{\theta})) \Big\}.
\end{equation*}
A similar one-step coordinate ascent update is then performed within the $j'_t$-th block, following the same procedure as in equation \eqref{eq:cavi}. The complete procedure is summarized in Algorithm \ref{alg:cavi}.

\begin{algorithm}[H]
	\caption{{\small $\mvb$ algorithm without local latent variables}}
	\label{alg:cavi}
	\hspace*{\algorithmicindent} \hspace{-0.7cm}   {\textbf{Input:} Data divided in $m$ subsets $\mbX_j=\{\vect{X}_i| i\in\mcH_j\}$, the number of iterations $T$.\\ \mbox{}} 	
	\begin{algorithmic}[1]
		\FOR{$t=1,...,T$}
			\STATE Find $j_t\in[m]$ such that 
			\begin{equation*}
				\elbo_{j_t}(G^{(t)}(\vect{\theta})) - \elbo_{j_t}(F^{(t)}(\vect{\theta})) = \med{1\leq j\leq m}\Big\{ \elbo_{j}(G^{(t)}(\vect{\theta})) - \elbo_{j}(F^{(t)}(\vect{\theta})) \Big\}.
			\end{equation*}
			\FOR{$l=1,...,p$}
			\STATE Apply coordinate accent
			\begin{equation*}
				f_{l}^{(t+1)}(\theta_l) \propto \exp\big\{\mbE_{-l,F^{(t)}}[\log \{p^m(\mbX_j|\vect{\theta})\pi(\vect{\theta})\}]\big\},\quad f^{(t+1)}(\vect{\theta})=\prod_{l=1}^pf_l^{(t+1)}(\theta_l).
			\end{equation*}
			\ENDFOR
			\STATE Find $j'_t\in[m]$ such that 
			\begin{equation*}
				\elbo_{j'_t}(G^{(t)}(\vect{\theta})) - \elbo_{j'_t}(F^{(t+1)}(\vect{\theta})) = \med{1\leq j\leq m}\Big\{ \elbo_{j}(G^{(t)}(\vect{\theta})) - \elbo_{j}(F^{(t+1)}(\vect{\theta})) \Big\}.
			\end{equation*}
			\FOR{$l=1,...,p$}
			\STATE Apply coordinate accent
			\begin{equation*}
				g_{l}^{(t+1)}(\theta_l) \propto \exp\big\{\mbE_{-l,G^{(t)}}[\log \{p^m(\mbX_j|\vect{\theta})\pi(\vect{\theta})\}]\big\},\quad g^{(t+1)}(\vect{\theta})=\prod_{l=1}^pg_l^{(t+1)}(\theta_l).
			\end{equation*}
			\ENDFOR
		\ENDFOR
	\end{algorithmic}
	 \textbf{Output:}  The final distribution $F^{(T)}(\vect{\theta})$.
\end{algorithm}

\begin{example}	\label{exp:bayes_lr}
	We consider the Bayesian linear regression model as presented in \cite{Drugowitsch.2019arxiv}. The model is specified as 
	\begin{equation}   \label{eq:blr}
		Y = \vect{X}^{\tp}\vect{\beta}+\epsilon,\quad\epsilon\sim\mcN(0,\sigma^2),
	\end{equation}
	where $Y$ denotes the responsevariable and $\vect{X}=(X_1,...,X_p)^{\tp}$ is the covariate vector of dimension $p$. The goal is to infer the regression coefficient vector $\vect{\beta}$. We assume a conjugate normal-inverse-gamma prior for the pair $\vect{\beta},\sigma^2$, given by
	\begin{align*}
		P(\vect{\beta},\sigma^2) =& \mcN(\vect{\beta}; \vect{0}, \alpha\sigma^2\mbI_p)\mathrm{IG}(\sigma^2;a_0/2,b_0/2)\\
			\propto& \sigma^{-(p+a_0+2)}\exp\Big[-\frac{1}{2\sigma^2}\Big\{\frac{|\vect{\beta}|_2^2}{\alpha}+b_0\Big\}\Big].
	\end{align*}
	In this setup, there are no latent variables, and the parameter vector is $\vect{\theta}=(\vect{\beta}^{\tp},\sigma^2)^{\tp}$. We restrict the variational distribution family to be
	\begin{equation*}
		Q_{\vect{\theta}} = \bigotimes_{l=1}^pQ_{\beta_l}\otimes Q_{\sigma^2} = \bigotimes_{l=1}^p\mcN(\beta_l;\mu_l,\sigma_l^2)\otimes\mathrm{IG}(\sigma^2;c/2,d/2),
	\end{equation*}
	Leveraging the methodology from \cite{huang_etal.2016arxiv}, we present the $\mvb$ algorithm for Bayesian linear regression in Algorithm 3, provided in the appendix.
\end{example}

\begin{remark}
	In \cite{minsker_etal.2014, minsker_etal.2017jmlr, padilla_etal.2025arXiv}, the authors propose aggregating local (variational) posteriors via the geometric median. However, these approaches suffer from computational intractability in several respects. First, these methods rely on various metrics defined over the space of probability distributions, such as the Wasserstein distance and RKHS induced norms. Computing such distances between distributions is generally non-trivial and computationally demanding. Second, the algorithms for computing the geometric median of distributions often require case-specific implementations. For example, \cite{minsker_etal.2017jmlr} advocate the use of Weiszfeld's algorithm for discrete distributions, while \cite{padilla_etal.2025arXiv} propose a linear programming-based approach tailored to Gaussian mixture models. Finally, these approaches lack the flexibility to be seamlessly integrated into the variational Bayes framework. In contrast, our proposed method offers a unified algorithmic framework that is broadly applicable across a wide range of variational Bayes problem, while also being computationally efficient.
\end{remark}


\subsection{$\mvb$ Algorithm with Local Latent Variables}	\label{sec:alg_withll}

In this section, we develop an algorithm for the $\mvb$ estimator in the presence of local latent variables. Our approach leverages the following key property of the min-max median estimator: 

\begin{proposition}	\label{prop:latent_putin}
	For any variational family $\mcQ$, the following equality holds:
	\begin{align*}
		&\argmin{F_{\vect{\theta}}\in\mcQ^{p}}\min_{F_{\mbS}\in\mcQ^{mn}}\max_{G\in\mcQ^{p+mn}}\Big[\med{1\leq j\leq m}\Big\{\elbo(G_{\vect{\theta}}\otimes G_{\mbS_j}\| P(\vect{\theta},\mbS_j,\mbX_j))\\
		&-\elbo(F_{\vect{\theta}}\otimes F_{\mbS_j}\| P(\vect{\theta},\mbS_j,\mbX_j))\Big\}\Big]\\
		=&\argmin{F_{\vect{\theta}}\in\mcQ^{p}}\max_{G_{\vect{\theta}}\in\mcQ^{p}}\Big[\med{1\leq j\leq m}\Big\{\max_{G_{\mbS_j}\in\mcQ^{n}}\elbo(G_{\vect{\theta}}\otimes G_{\mbS_j}\| P(\vect{\theta},\mbS_j,\mbX_j))\\
		&-\max_{F_{\mbS_j}\in\mcQ^{n}}\elbo(F_{\vect{\theta}}\otimes F_{\mbS_j}\| P(\vect{\theta},\mbS_j,\mbX_j))\Big\}\Big].
	\end{align*}
\end{proposition}

Proposition \ref{prop:latent_putin} demonstrates that the minimization (maximization) over the local latent variables $\{\mbS_j\}_{j=1}^m$ can be interchanged with the median operation. This property significantly simplifies both the theoretical analysis and the development of efficient algorithms. For notational convenience, we define
\begin{equation*}
	\elbo_j(F) = \elbo(F_{\vect{\theta}}\otimes F_{\mbS_j}\|P(\vect{\theta},\mbS_j,\mbX_j)),
\end{equation*}
At iteration $t$, given the current distribution $F^{(t)}$ and $G^{(t)}$, we select the index $j_t$ such that
\begin{equation*}
	\elbo_{j_t}(G^{(t)}(\vect{\theta})) - \elbo_{j_t}(F^{(t)}(\vect{\theta})) = \med{1\leq j\leq m}\Big\{ \elbo_{j}(G^{(t)}(\vect{\theta})) - \elbo_{j}(F^{(t)}(\vect{\theta})) \Big\}.
\end{equation*}
We then perform a single coordinate ascent step within the $j_t$-th block. More specifically, the update for the $l$-th coordinate of $F^{(t)}_{\vect{\theta}}$ is given by 
\begin{equation*}	
	f_{l}^{(t+1)}(\theta_l) \propto \exp\big\{\mbE_{-l,F^{(t)}}[\log \{p(\mbX_j|\vect{\theta},\mbS_j)p(\mbS_j|\vect{\theta})\pi(\vect{\theta})\}]\big\},
\end{equation*}
from which we obtain the updated distribution $F^{(t+1)}_{\vect{\theta}}$ by $f^{(t+1)}_{\vect{\theta}} = \prod_{l=1}^pf_l^{(t+1)}(\theta_l)$. Next, we update the distributions of all local latent variable $S_i$, $1\leq i\leq N$. Leveraging Proposition \ref{prop:latent_putin}, the local $\elbo$ can be maximized within the median, yielding 
\begin{equation*}
	f_{l}^{(t+1)}(S_i) \propto \exp\big\{\mbE_{-l,F^{(t)}}[\log \{p(\vect{X}_i|\vect{\theta},S_i)p(S_i|\vect{\theta})\pi(\vect{\theta})\}]\big\}.
\end{equation*}
This provides the full updated distribution $F^{(t+1)}$. The same procedure is applied symmetrically to update $G^{(t)}$. The complete algorithm is summarized in Algorithm \ref{alg:cavi_ll}.

\begin{algorithm}[H]
	\caption{{\small $\mvb$ algorithm with local latent variables}}
	\label{alg:cavi_ll}
	\hspace*{\algorithmicindent} \hspace{-0.7cm}   {\textbf{Input:} Data divided in $m$ subsets $\mbX_j=\{\vect{X}_i| i\in\mcH_j\}$, the number of iterations $T$.\\ \mbox{}} 	
	\begin{algorithmic}[1]
		\FOR{$t=1,...,T$}
			\STATE Find $j_t\in[m]$ such that 
			\begin{equation*}
				\elbo_{j_t}(G^{(t)}) - \elbo_{j_t}(F^{(t)}) = \med{1\leq j\leq m}\Big\{ \elbo_{j}(G^{(t)}) - \elbo_{j}(F^{(t)}) \Big\}.
			\end{equation*}
			\FOR{$l=1,...,p$}
			\STATE Apply coordinate accent
			\begin{equation*}
				f_{l}^{(t+1)}(\theta_l) \propto \exp\big\{\mbE_{-l,F^{(t)}}[\log \{p(\mbX_j|\vect{\theta},\mbS_j)p(\mbS_j|\vect{\theta})\pi(\vect{\theta})\}]\big\},\quad f_{\vect{\theta}}^{(t+1)}(\vect{\theta})=\prod_{l=1}^pf_l^{(t+1)}(\theta_l).
			\end{equation*}
			\ENDFOR
			\FOR{$i=1,...,N$}
			\STATE Apply coordinate accent
			\begin{equation*}
				f_{i}^{(t+1)}(S_i) \propto \exp\big\{\mbE_{-i,F^{(t)}}[\log \{p(\vect{X}_i|\vect{\theta},S_i)p(S_i|\vect{\theta})\pi(\vect{\theta})\}]\big\},\quad f^{(t+1)}=f^{(t+1)}_{\vect{\theta}}\prod_{i=1}^Nf_i^{(t+1)}(S_i).
			\end{equation*}
			\ENDFOR
			\STATE Find $j'_t\in[m]$ such that 
			\begin{equation*}
				\elbo_{j'_t}(G^{(t)}) - \elbo_{j'_t}(F^{(t+1)}) = \med{1\leq j\leq m}\Big\{ \elbo_{j}(G^{(t)}) - \elbo_{j}(F^{(t+1)}) \Big\}.
			\end{equation*}
			\FOR{$l=1,...,p$}
			\STATE Apply coordinate accent
			\begin{equation*}
				g_{l}^{(t+1)}(\theta_l) \propto \exp\big\{\mbE_{-l,G^{(t)}}[\log \{p(\mbX_j|\vect{\theta},\mbS_j)p(\mbS_j|\vect{\theta})\pi(\vect{\theta})\}]\big\},\quad g_{\vect{\theta}}^{(t+1)}(\vect{\theta})=\prod_{l=1}^pg_l^{(t+1)}(\theta_l).
			\end{equation*}
			\ENDFOR
			\FOR{$i=1,...,N$}
			\STATE Apply coordinate accent
			\begin{equation*}
				g_{i}^{(t+1)}(S_i) \propto \exp\big\{\mbE_{-i,G^{(t)}}[\log \{p(\vect{X}_i|\vect{\theta},S_i)p(S_i|\vect{\theta})\pi(\vect{\theta})\}]\big\},\quad  g^{(t+1)}=g^{(t+1)}_{\vect{\theta}}\prod_{i=1}^Ng_i^{(t+1)}(S_i).
			\end{equation*}
			\ENDFOR
		\ENDFOR
		\STATE Rescale distribution $F^{(T)}(\vect{\theta})$ by \eqref{eq:ftilde_rescale}.
	\end{algorithmic}
	 \textbf{Output:}  The final rescaled distribution $\hat{F}^{(T)}(\vect{\theta})$.
\end{algorithm}

\begin{example} \label{exp:gmm}
	Consider a Gaussian mixture model consisting of $K$ univariate Gaussian components, each with unit variance and means given by the vector $\vect{\theta} = (\theta_1,...,\theta_K)^{\tp}$. For each observation $\vect{X}_i, i =1,...,N$, a cluster assignment $S_i$ is first drawn from a categorical distribution over the set $\{1,...,K\}$. Conditioned on $S_i$, the observation $\vect{X}_i$ is then generated from a Gaussian distribution $\mcN(\theta_{S_i},1)$. Formally, the model is specified as
	\begin{align*}
		&\theta_l\sim \mcN(0,\sigma_0^2),\quad && l =1,...,K;\\
		&S_i \sim\mathrm{Categorical}(1/K,...,1/K),\quad&& i=1,...,N;\\
		&\vect{X}_i|S_i,\vect{\theta}\sim\mcN(\theta_{S_i},1),\quad&&i=1,...,N.
	\end{align*}
	We constrain the variational family to densities of the form
	\begin{equation*}
		q(\vect{\theta},\mbS) = \prod_{l=1}^pq(\theta_l;m_l,s_l^2)\prod_{i=1}^Nq(S_i;\phi_i),
	\end{equation*}
	where $q(\theta_l;m_l,s_l^2)$ denotes the density of a normal distribution $\mcN(m_l,s_l^2)$, and $q(S_i;\phi_i)$ is a categorical distribution over cluster assignments for the $i$-th observation, with probability vector $\phi_i$. Following \cite{Blei_etal.2017jasa}, the mean-field variational Bayes algorithm for the Gaussian mixture model is presented in Algorithm 4 in the appendix.
\end{example}


\section{Simulation Study}	\label{sec:sim}

In this section, we present numerical experiments to evaluate the performance of our proposed methods. We focus on two models: Bayesian linear regression, which does not involve local latent variables, and the Gaussian mixture model, which includes local latent variables. In all simulations, we vary the local sample size $n$ over the set $\{100,200,400,600,1000,1500,2000\}$ and the number of data subsets $m$ over $\{20,30,40\}$. Each experimental setting is repeated $100$ times to ensure statistical reliability.


\subsection{Bayesian Linear Regression}	\label{sec:sim_blr}

In the first set of experiments, we evaluate our proposed method using the Bayesian linear regression model described in Example \ref{exp:bayes_lr}. Data are generated according to model \eqref{eq:blr} with dimensionality $p=6$. We set the true regression coefficient $\vect{\beta} = (2.0, -1.0, 0.5, 0.0, 1.5, -0.5)^\top$, and the noise variance to $\sigma=1$. Covariate vectors $\vect{X}$ are sampled from $\mcN(\vect{0},\mbI_p)$. To assess robustness, a fraction $\alpha_n$ of subsets are intentionally corrupted: their response values $Y$ are drawn instead from $\mathcal{N}(10, 1)$, introducing outliers into the dataset.

\paragraph{Effect of $n$ and $m$}

We begin by examining the impact of the local sampled size $n$ and the number of subsets $m$, fixing the corruption fraction at $\alpha_n=0.05$. In addition to our proposed $\mvb$ method, we consider two competing approaches:
\begin{itemize}
    \item Pooled VB: The corrupted data are removed, and the standard variational Bayes method is applied to the remaining clean data. This serves as a benchmark for comparison;
    \item WASP: The method of \cite{srivastava_etal.2015aistats}, which aggregates local posterior distributions by computing their Wasserstein barycenter.
\end{itemize}

\begin{figure}[htbp]
    \centering

    \begin{subfigure}{1.0\textwidth}
        \centering
        \includegraphics[width=\linewidth]{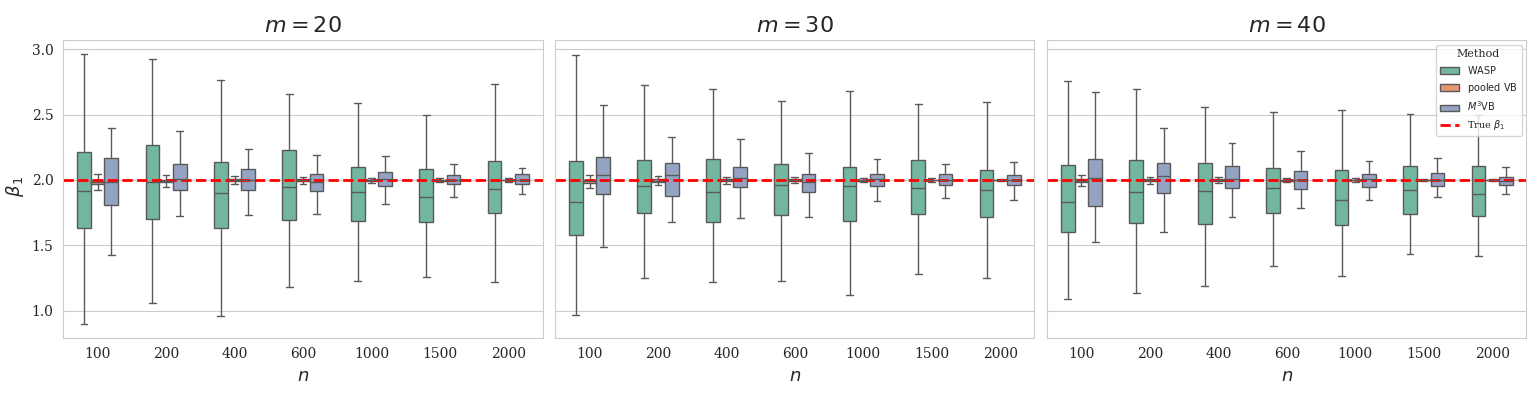}
    \end{subfigure}%

    \begin{subfigure}{1.0\textwidth}
        \centering
        \includegraphics[width=\linewidth]{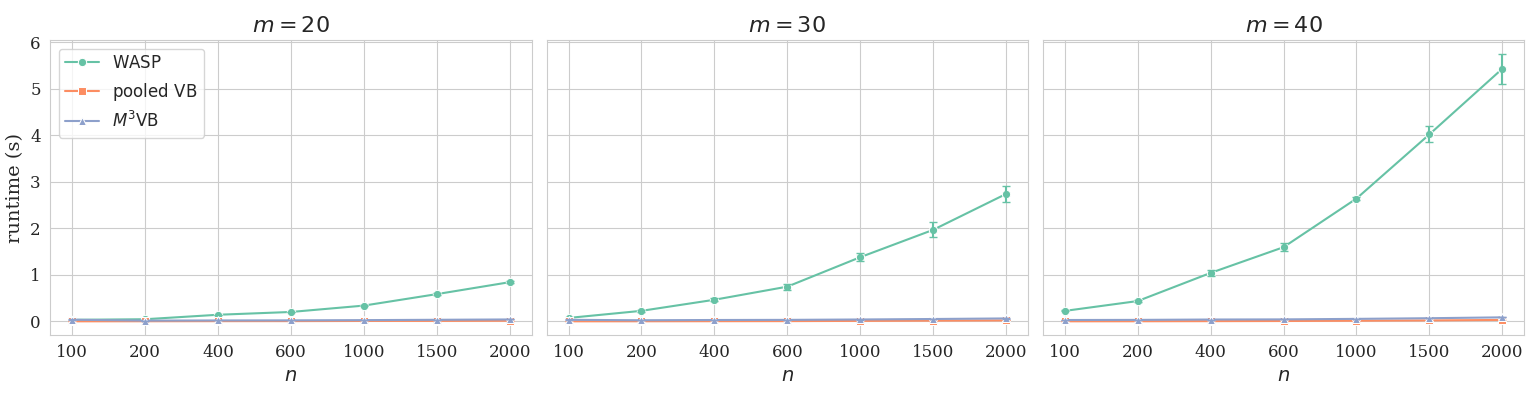}
    \end{subfigure}
    \hfill

    \caption{Boxplots of posterior estimate for $\beta_1$ (top row) and line plots of computation times (bottom row) under Bayesian linear regression across different methods. The number of subsets $m$ varies over $\{20,30,40\}$, and the number of local samples $n$ ranges over $\{100,200,400,600,1000,1500,2000\}$. The corruption rate is $\alpha_n=0.05$.}
    \label{fig:lr_mn}
\end{figure}

It is important to note that the WASP method is generally not robust to outliers. \cite{padilla_etal.2025arXiv} proposed a robust VB approach based on computing the geometric median of distributions; however, due to its high computational cost, we do not include it in our experiments. We report boxplots of the distribution mean for the first parameter $\beta_1$, and compare the running time of the methods. The results are presented in Figure \ref{fig:lr_mn}.

From Figure \ref{fig:lr_mn}, it is evident that our $\mvb$ method exhibits robustness to data corruption while achieving computational efficiency comparable to that of pooled VB. The variability in the boxplots decreases as the local sample size $n$ increases. In contrast, the WASP method is more sensitive to outliers and incurs a higher computational cost.

\paragraph{Compare with Two-Step Approach}

\begin{figure}[htbp]
    \centering

    \begin{subfigure}{1.0\textwidth}
        \centering
        \includegraphics[width=\linewidth]{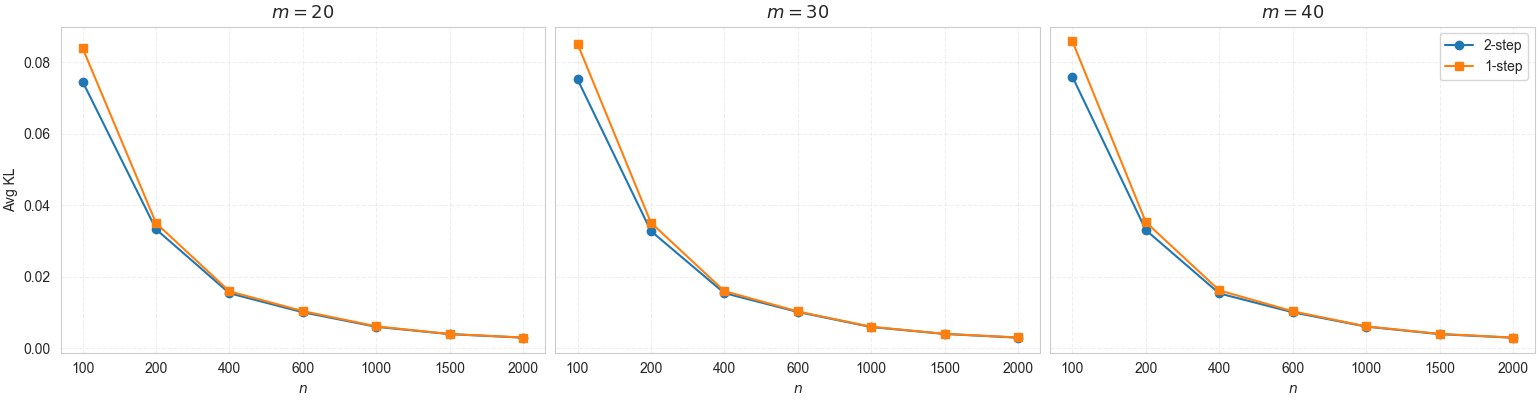}
    \end{subfigure}%

    \begin{subfigure}{1.0\textwidth}
        \centering
        \includegraphics[width=\linewidth]{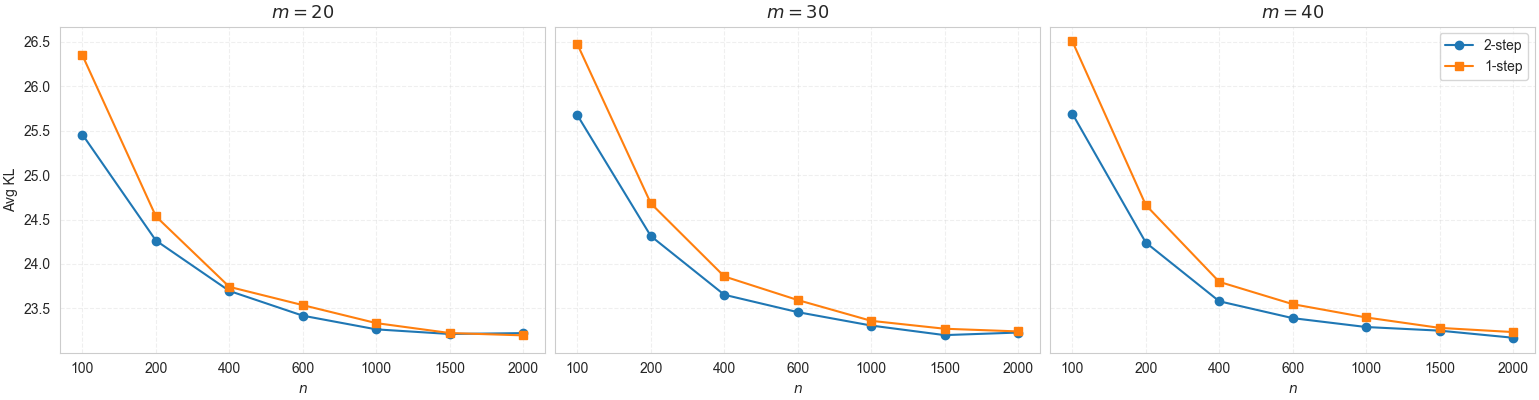}
    \end{subfigure}
    \hfill

    \caption{Line plots of the KL divergence from $Q^*(\vect{\mu}_{\hat{F}_{\vect{\theta}}})$ under Bayesian linear regression for two methods. The corruption rate is $\alpha_n=0$ in the top row and $\alpha_n=0.1$ in the bottom row. The number of subsets $m$ varies over $\{20,30,40\}$, and the number of local labeled samples $n$ ranges across $\{100,200,400,600,1000,1500,2000\}$.}
    \label{fig:lr_12step}
\end{figure}

In Section \ref{sec:med_elbo}, we introduce the two-step $\mvb$ method specifically tailored for models incorporating local latent variables. It is important to note that this two-step procedure can be seamlessly extended to contexts without local latent variables. Analogous to the results in Corollary \ref{cor:BvM_ll}, we can prove that the two-step approach yields a smaller KL divergence than the one-step procedure. In our experiments, we assess the performance of these two methods under corruption rates $\alpha_n \in \{0,\,0.1\}$, and present line plots of the KL divergence from the reference distribution $Q^*(\vect{\mu}_{\hat{F}_{\vect{\theta}}})$, as defined in Corollary~\ref{cor:BvM_ll}. The outcomes are depicted in Figure \ref{fig:lr_12step}.

From Figure \ref{fig:lr_12step}, it is evident that the two-step $\mvb$ method achieves consistently lower KL divergence compared to the one-step approach, aligning with the theoretical expectations from Corollary \ref{cor:BvM_ll}. However, the magnitude of improvement is modest. Notably, when the corruption rate is set to $\alpha_n=0.1$, the KL divergence exceeds 20, albeit it decreases as the sample size $n$ increase. This observation suggests that there may be room for further refinement in our theoretical analysis.

\paragraph{Compare with Direct Median Approach}

\begin{figure}[htbp]
    \centering

    \begin{subfigure}{0.48\textwidth}
        \centering
        \includegraphics[width=\linewidth]{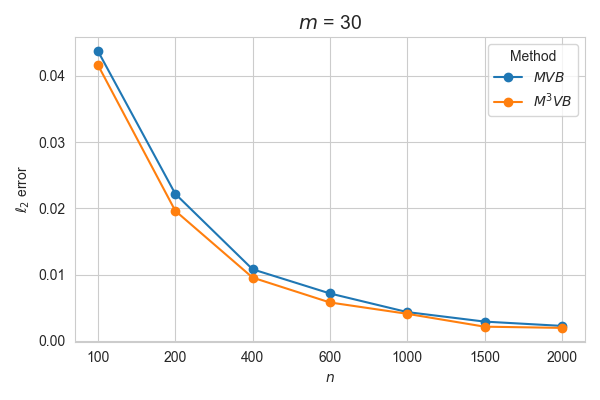}
    \end{subfigure}\hfill
    \begin{subfigure}{0.48\textwidth}
        \centering
        \includegraphics[width=\linewidth]{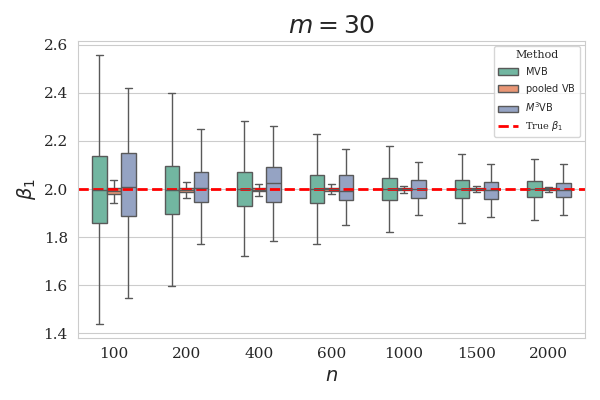}
    \end{subfigure}
    \hfill

    \caption{Line plot of the estimation error (left) and boxplots of the parameter $\beta_1$ (right) for various methods applied to Bayesian linear regression. The number of subsets is fixed at $m=30$, while the number of local labeled samples $n$ varies over $\{100,200,400,600,1000,1500,2000\}$.}
    \label{fig:lr_mvb}
\end{figure}

In Remark \ref{rem:mvb_mean}, we discussed that the direct median-based approach defined in \eqref{eq:naive_KLVB_median} may exhibit a slower convergence rate in the distribution mean. To further investigate this, we fix the number of subsets at $m=30$, and compare our proposed min-max method with the direct median-based method ( denoted as $\mathrm{MVB}$). We present both line plots of the estimation error and boxplots of the first coordinate $\beta_1$. Results are displayed in Figure \ref{fig:lr_mvb}.

From Figure \ref{fig:lr_mvb}, it is apparent that our min-max formulation consistently yields lower $\ell_2$ estimation errors than the direct median-based approach. Additionally, the boxplots corresponding to $\mvb$ are invariably narrower than those for $\mathrm{MVB}$. These findings provide empirical support for the theoretical benefits of the min-max formulation.


\subsection{Gaussian Mixture Model}	\label{sec:sim_gmm}

In the second set of experiments, we examine the performance of our method on the Gaussian mixture model described in Example \ref{exp:gmm}, using $K=3$ components, each a univariate Gaussian with unit variance. The true component means are set to $\vect{\theta} = (-3.0,\, 0.0,\, 3.0)^{\top}$ with equal mixing proportions $(1/3,\, 1/3,\, 1/3)^{\tp}$. To assess robustness, we introduce corruption in a fraction $\alpha_n=0.05$ of the data, replacing these observations with draws from $\mcN(0,5)$.

\paragraph{Effect of $n$ and $m$}

\begin{figure}[htbp]
    \centering

    \begin{subfigure}{1.0\textwidth}
        \centering
        \includegraphics[width=\linewidth]{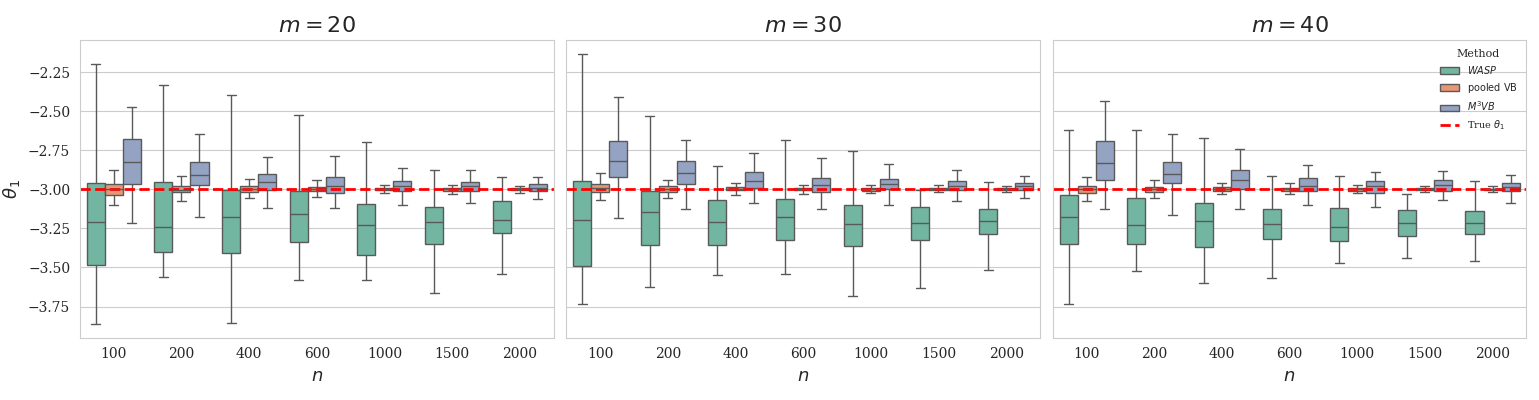}
    \end{subfigure}%

    \begin{subfigure}{1.0\textwidth}
        \centering
        \includegraphics[width=\linewidth]{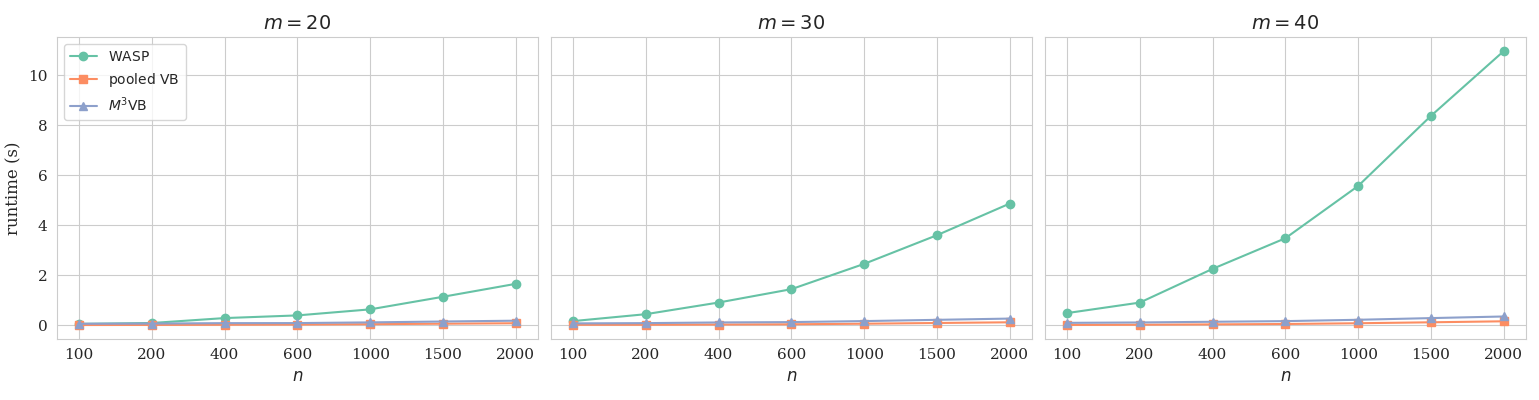}
    \end{subfigure}
    \hfill

    \caption{Boxplots of posterior estimate for $\theta_1$ (top row) and line plots of computation times (bottom row) under Gaussian mixture model across different methods. The number of subsets $m$ varies over $\{20,30,40\}$, and the number of local samples $n$ ranges over $\{100,200,400,600,1000,1500,2000\}$. The corruption rate is $\alpha_n=0.05$.}
    \label{fig:gmm_mn}
\end{figure}

Analogous to the previous section, we evaluate the effects of sample size $n$ and number of subsets $m$, comparing our method against Pooled VB and WASP method. We present boxplots of the first coordinate $\theta_1$ and the computational runtime for each method in Figure \ref{fig:gmm_mn}. The results demonstrate that our proposed approach maintains both robustness to corruption and computational efficiency.

\paragraph{Compare with One-Step Approach}

\begin{figure}[htbp]
    \centering

    \begin{subfigure}{1.0\textwidth}
        \centering
        \includegraphics[width=\linewidth]{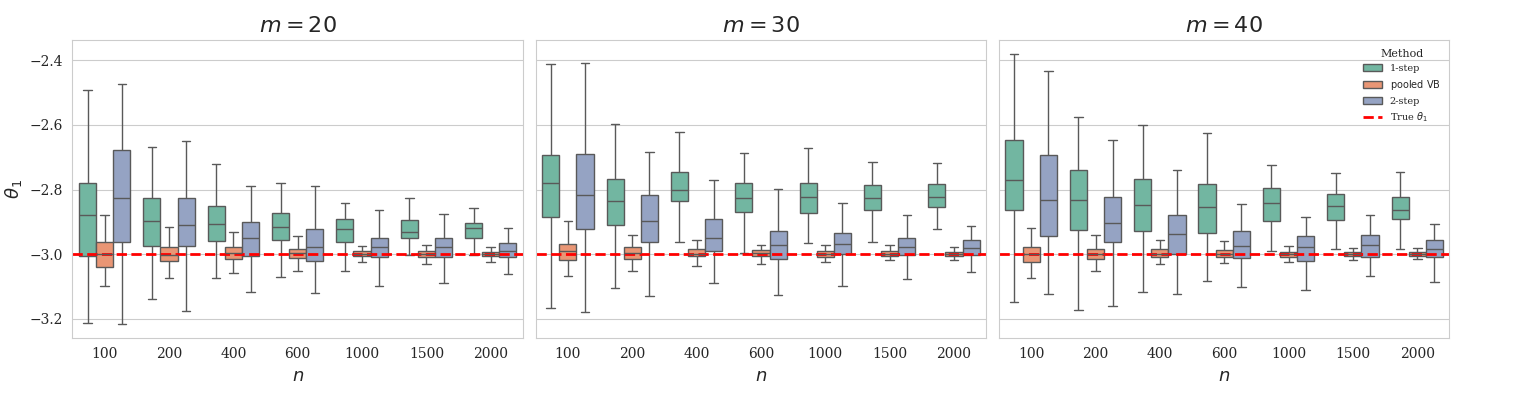}
    \end{subfigure}%

    \caption{Boxplots of the parameter $\theta_1$ across different methods under Gaussian mixture model. The number of subsets $m$ varies among $\{20,30,40\}$, and the number of local labeled samples $n$ spans $\{100,200,400,600,1000,1500,2000\}$.}
    \label{fig:gmm_12step}
\end{figure}

In Section \ref{sec:med_elbo}, we noted that the one-step $\mvb$ approach introduces a non-negligible bias in the distribution mean. To empirically validate this observation, we conduct an experiment comparing the single-step $\mvb$ method with its two-step variant and the Pooled VB benchmark. We present boxplots for the first coordinate $\theta_1$, as shown in Figure \ref{fig:gmm_12step}. The figure clearly demonstrates that the one-step approach produces a pronounced bias in the distribution mean, which persists even as the local sample size $n$ increases. In contrast, the two-step $\mvb$ procedure significantly mitigates this bias, affirming the theoretical insights discussed in Section \ref{sec:med_elbo}.

\paragraph{Compare with Direct Median Approach}

\begin{figure}[htbp]
    \centering

    \begin{subfigure}{0.48\textwidth}
        \centering
        \includegraphics[width=\linewidth]{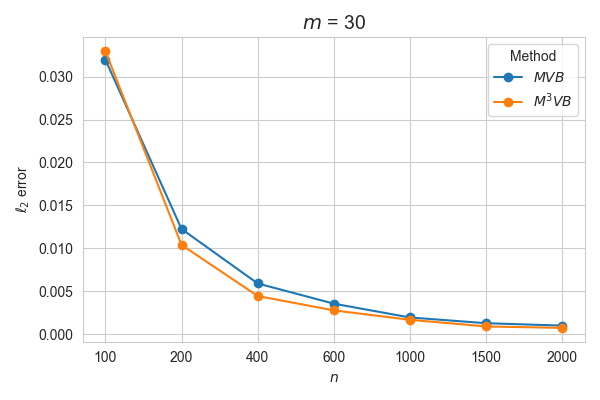}
    \end{subfigure}\hfill
    \begin{subfigure}{0.48\textwidth}
        \centering
        \includegraphics[width=\linewidth]{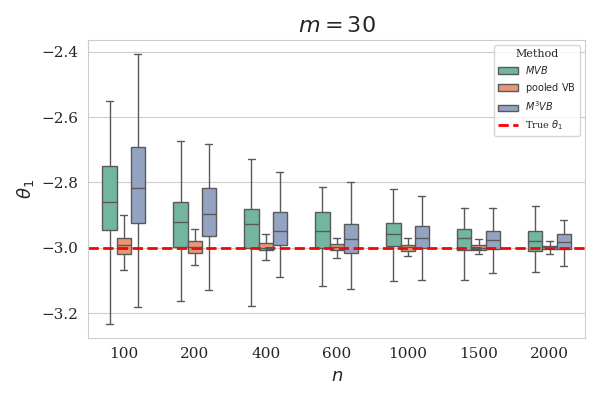}
    \end{subfigure}
    \hfill

    \caption{Line plot of the estimation error (left) and boxplots of the parameter $\theta_1$ (right) for various methods applied to Gaussian mixture model. The number of subsets is fixed at $m=30$, while the number of local labeled samples $n$ varies over $\{100,200,400,600,1000,1500,2000\}$.}
    \label{fig:gmm_mvb}
\end{figure}

In the final experiment, we fix the number of subsets at $m=30$, and compare our method to the direct median-based approach (denoted as $\mathrm{MVB}$). We visualize the results using both a line plot of the $\ell_2$ estimation error and a boxplot of the first parameter coordinate $\theta_1$. These results are displayed in Figure \ref{fig:gmm_mvb}. The empirical evidence clearly indicates that our min-max approach consistently outperforms the direct median method in terms of $\ell_2$ error, and the associated boxplots for $\mvb$ exhibit substantially narrower variability than those for $\mathrm{MVB}$.

\section{Concluding Remarks}	\label{sec:conclude}

In conclusion, this paper propose a robust variational Bayes method based on a min-max median aggregation framework. To ensure statistical efficiency comparable to the full sample posterior, we advocate aggregating the $m$-powered local posteriors in settings without local latent variables. However, in scenarios involving local latent variables, we observe that the $m$-powered local posterior leads to a biased estimator. To address this issue, we introduce an aggregate-and-rescale procedure, a two-stage approach designed to correct the bias. Theoretically, we establish a non-asymptotic Bernstein-von Mises theorem for the proposed $\mvb$ estimator, demonstrating that the two-stage estimator achieves a smaller approximation error compared to directly applying $\mvb$ to the $m$-powered local posteriors. Additionally, we provide a refined analysis of the statistical convergence rate of the posterior mean, improving upon existing median-of-means results.

Several avenues merit further investigation. First, our theoretical results show that the $\mvb$ posterior mean attains a near-optimal statistical rate only under a restrictive condition on the number of subsets, specifically $m=O(\sqrt{n})$. This limitation arise because the aggregation procedure reduces variance but not eliminate local bias, a phenomenon also observed in distributed learning contexts \citep{zhang_etal.2013jmlr, jordan_etal.2019}. Developing alternative methods that effectively mitigate local bias remains an open and important problem. Second, while our aggregate-and-rescale approach yields a smaller approximation error, the rescaling operation defined in \eqref{eq:ftilde_rescale} is generally intractable, as the variational posterior mean may not be easily computed. Designing a one-step procedure that achieves comparable efficiency for variational Bayes with local latent variables is a promising direction. Lastly, our method addresses a min-max median optimization problem, which is inherently non-convex and non-smooth. Although the algorithm proposed in Section \ref{sec:alg} is simple and practically effective, it lacks theoretical guarantees for convergence to near-optimal solutions. Constructing algorithms with provable consistency guarantees for the min-max estimator remains an important challenge.


\bibliographystyle{asa}
\bibliography{M3VB_arXiv}

\end{document}